\documentclass[11pt,aps,prd,notitlepage,
nofootinbib,
superscriptaddress,showkeys,showpacs]{revtex4-2}

\usepackage{enumitem}
\usepackage{amsfonts, amsmath, amssymb, amsthm}
\usepackage{color}
\usepackage{stmaryrd}
\usepackage{graphicx, array}
\usepackage{bbm}
\usepackage{hyperref}
\usepackage{mdframed}
\usepackage[all]{xy}
\usepackage{changepage}
\usepackage{titlesec}

\newcommand{\be}{\begin{equation}}
\newcommand{\ee}{\end{equation}}
\newcommand{\bea}{\begin{eqnarray}}
\newcommand{\eea}{\end{eqnarray}}
\renewcommand\[{\begin{equation}}
\renewcommand\]{\end{equation}}

\def\d{\mathrm{d}}
\def\D{\mathrm{D}}
\def\pd{\partial}
\def\e{\textrm e}
\def\i{\imath}

\newcommand{\tr}{\mathrm{Tr}}
\newcommand{\U}{\textrm{U}} 
\newcommand{\vd}{\Delta}	
\newcommand{\amp}{\mathcal{A}}

\newcommand{\dx}{\frac{\d}{\d x}}
\newcommand{\dxi}[1]{\frac{\d^{#1}}{\d x^{#1}}}
\newcommand{\dy}{\frac{\d}{\d y}}
\newcommand{\dyi}[1]{\frac{\d^{#1}}{\d y^{#1}}}
\newcommand{\x}{\mathbf{x}}
\newcommand{\y}{\mathbf{y}}
\newcommand{\bp}{\mathbf{p}}
\newcommand{\cZ}{\mathcal{Z}}
\newcommand{\ba}{\mathbf{a}}
\newcommand{\bn}{\mathbf{n}}
\newcommand{\bb}{\mathbf{b}}
\newcommand{\bfm}{{\mathbf{m}}}
\newcommand{\bH}{\mathbb{H}}
\newcommand{\bR}{\mathbb{R}}

\renewcommand{\c}{\alpha} 

\newcommand{\diagblock}[2]{\begin{array}{c} #1 \\ \eqref{#2} \end{array}}

\makeatletter 
\renewcommand\Large{\@setfontsize\Large{12pt}{18}}
\renewcommand\large{\@setfontsize\large{12pt}{18}}
\makeatother

\renewcommand\thesection{\Roman{section}}
\renewcommand\thesubsection{\thesection.\,\arabic{subsection}}
\renewcommand\thesubsubsection{(\alph{subsubsection})}

\makeatletter
\renewcommand{\p@section}{}
\renewcommand{\p@subsection}{}
\renewcommand{\p@subsubsection}{\thesubsection.\,}
\makeatother

\titleformat{\section}{\Large\bfseries\filcenter}{\thesection. }{0em}{}
\titleformat{\subsection}{\large\bfseries\filcenter}{\thesubsection. }{0em}{}
\titleformat{\subsubsection}[runin]{\bfseries}{\thesubsubsection}{0.5em}{}[.]
\titlespacing{\subsubsection}{0pt}{10pt}{1em}

\makeatletter
\newcommand*\bigcdot{\mathpalette\bigcdot@{.5}}
\newcommand*\bigcdot@[2]{\mathbin{\vcenter{\hbox{\scalebox{#2}{$\m@th#1\bullet$}}}}}
\makeatother

\begin{document}

\title{One-matrix differential reformulation of two-matrix models}

\author{Joren Brunekreef}
\email{jorenb@gmail.com}
\author{Luca Lionni}
\email{luca.lionni@ru.nl}
\affiliation{IMAPP, Radboud University, Nijmegen, The Netherlands.}

\author{Johannes Th\"urigen}
\email{johannes.thuerigen@uni-muenster.de}
\affiliation{Mathematisches Institut der Westf\"alischen Wilhelms-Universit\"at M\"unster\\
Einsteinstr.\,62, D-48419 M\"unster, EU}
\affiliation{Institut f\"ur Mathematik/Institut f\"ur Physik der Humboldt-Universit\"at zu Berlin\\
Unter den Linden 6, D-10099 Berlin, EU}

\date{\today}

\begin{abstract} 
Differential reformulations of field theories are often used for explicit computations. 
We derive a one-matrix differential formulation of two-matrix models, with the help of which it is possible to diagonalize the one- and two-matrix models using a formula by Itzykson and Zuber that allows diagonalizing differential operators with respect to matrix elements of Hermitian matrices. 
We detail the equivalence between the expressions obtained by diagonalizing the partition function in differential or integral formulation, which is not manifest at first glance. For one-matrix models, this requires transforming certain derivatives to variables.
In the case of two-matrix models, the same computation leads to a new determinant formulation of the partition function, and we discuss potential applications to new orthogonal polynomials methods.
\end{abstract}

\maketitle

\section{Introduction}
Two-matrix models have been widely studied in the literature since their introduction by Itzykson and Zuber \cite{Itzykson:1980hi}, see for instance \cite{ Mehta:1981jm, 1997NuPhB.506..633E, Eynard:2003hs,Bertola:2003cn,Bertola:2007is, Duits:2012vg}. Their Feynman expansions count ribbon graphs with two kinds of vertices.
They match, for different choices of potentials, the partition functions of some statistical models such as the Ising model \cite{KAZAKOV-Ising}, colored triangulations \cite{ColoringRandomTriang}, hard particles \cite{Bouttier_2002} etc. on 
random surfaces (thereby relating to 2D quantum gravity). 

In quantum field theory, differential formulations such as
\be
\label{eq:central-identity}
\begin{split}
\frac{\cZ}{\cZ_0} &= \frac{1}{\cZ_0} \int \D\phi \, \e^{- \int \d^d x \, \d^d y\,\frac 1 2 \phi(x)  K(x,y)  \phi(y) - \int \d^d x \, V\left(\phi(x)\right)}\\ &= \left[\e^{- \int \d^d w \, V\left(\frac{\pd}{\pd {J(w)}}\right) } \e^{\int \d^d x\, \d^d y\,  J(x) K^{-1}(x,y)  J(y)}\right]_{J=0}\,,
\end{split}
\ee
are widely used (Zee coins it the ``central identity of quantum field theory'' in his famous book \cite{Zee:book}).
Here the $\phi$ are fields on a $D$-dimensional spacetime, $\int\D\phi$ is a functional integration, the interaction potential $V$ is a polynomial in the fields, $K$ is the propagator, and $\cZ_0$ is a normalization.   
Such formulations allow expressing Gaussian expectation values and Wick's formula, the key ingredient of perturbative expansions, as well-defined {algebraic} expressions that do not suffer the potential divergences of the combinatorially equivalent integrals (see for instance \cite{Salmhofer:book, Bauerschmidt:book, GurauKraj, Gurau:2014}). They moreover provide a practical tool for computations of Feynman diagrams. 

In particular, the differential formulation provides well-defined expressions for the Gaussian expectation values of \emph{two-matrix models}, while the integrals are not well-defined and only understood formally through Wick's theorem.

Remarkably, as we will show in the paper, two-matrix models can be expressed in terms of a single matrix in a differential formulation:
the two matrices necessary to control the interaction between the two kinds of vertices (or equivalently impose bipartiteness of the ribbon graphs) 
are no longer needed in the differential formulations. Since the necessity of using two matrices complicates the resolution of two-matrix models, the existence of a one-matrix differential formulation raises the question whether this allows for a simpler resolution.

With the help of the one-matrix differential reformulation, it is possible to diagonalize the two-matrix models directly in the differential formulation.
The steps leading to the expression of the partition function as a Slater determinant and thereby to the resolution by biorthogonal polynomials involve elegant and interesting equations whose equivalence is not always manifest. 
We detail their relations and take the opportunity to review and clarify some aspects of the differential formulation that can be found in the matrix-model literature, as well as the formula needed to diagonalize the differential operators with respect to matrix elements for Hermitian matrices, implicitly shown by Itzykson and Zuber in \cite{Itzykson:1980hi} and stated by Zuber in~\cite{Zuber:2008kn}. These computations do not seem to indicate any simplification in the resolution of two-matrix models in differential formulation as initially hoped: 
the role played by the two matrices is now played by the matrix and the derivatives with respect to the matrix, leading to the same resolution by biorthogonal polynomials.

On the other hand, the analysis of the equivalence between the different formulas for one-matrix models does provide new insights on two-matrix models. Indeed, the proof of equivalence between two of the diagonalized differential formulations of the one-matrix models involves transforming certain derivatives to variables in the Slater determinant formulation. The computation is not trivial and is detailed in the last section. The same computation can 
be applied with little modifications to two-matrix models, and leads to a new determinant formulation of their partition functions. 
Building on this, we comment on a potential new resolution using orthogonal polynomials instead of biorthogonal polynomials.

In order to show that the different expressions of the partition functions indeed coincide, we verify that all the prefactors and normalizations match. These constants are often omitted in the literature as they are sometimes tedious to compute and do not contribute to the critical behaviors or the combinatorial interpretations.
An advantage of the computations in differential formulations is that they most often do not need an overall normalization, as seen for instance in \eqref{eq:central-identity}.

Differential formulations are often used in the literature with little or without justifications or references, and in addition to the developments presented here, the present paper is also intended to provide a text of reference by giving an exhaustive review in the context of random matrix models.

\

For the benefit of the reader, we first present a diagrammatic overview of all the different representations of the one- and two-matrix models that we consider in this article, with lines connecting two representations when we have a direct computation for translating between them. The expressions are shown in a schematic way, omitting normalization terms, traces, factors of $N$, and the like. The full form of each expression can be found in the main text of the article. Function arguments and integration variables are omitted where possible. If confusion may arise, the arguments of the potential functions $V$ (for one-matrix models) and $V_1, V_2$ (for two-matrix models) are indicated by superscripts. The action is understood to be $S=N \tr(\frac{1}{2}M^2-V(M))$ for one-matrix models, and $S=N \tr(AB-V_1(A)-V_2(B))$ for two-matrix models. Elements of the matrices appearing inside the determinants are labeled  $i,j$. In the external expressions \eqref{eq:general-new-expression-p-s-diff} and \eqref{eq:general-new-expression-p-s-int}, the $r, s, p$ are related to the $i$ of \eqref{eq:Slater-diff} and \eqref{eq:slater-2-mat-usual} as $i=(p-1)r+s$ for a polynomial potential $V_1(x) = \sum_k^p \frac{\alpha_k}{k} x^k$ of degree $p$. 

The structure is the same in both diagrams. In the bottom layer, we have all the integral representations of the matrix model under consideration. In the top layer, we show the corresponding expressions in differential form. $\vd$ denotes the Vandermonde determinant. The front layers gather the expressions obtained after diagonalizing in integral form, while the back layers gather the expressions obtained after diagonalizing in differential form. 

Several lines connecting parts within a diagram share common features, indicated by a common greek letter. The lines $\alpha$ symbolize the direct  equivalence between integral and differential formulations. Lines $\beta$ symbolize the equivalence between the one- and two-matrix differential formulations. Lines  $\delta$ involve a diagonalization procedure in integral formulation, leading to the appearance of Vandermonde determinants. Similarly, lines $\eta$ refer to a diagonalization in the differential formulation. When moving from left to right along lines marked $\phi$, the common strategy is to expand the Vandermonde determinants, thereby moving the integration or differentiation inside the matrix elements of the determinant.
The appearance of $\kappa$ indicates that a heat kernel method was used to transform between an integral and differential expression. The steps marked $\sigma$ absorb the $1/\vd$ prefactor by extracting the symmetric part of the determinant.
Finally, for vertical lines labeled $\lambda$, it is shown how to transform some of the derivatives to variables and vice-versa, while $\tilde \lambda$ corresponds to the equivalent computation in integral formulation. These are the computations leading to our newly found determinantal expressions for two-matrix models, that could potentially lead to a new resolution in terms of orthogonal polynomials.

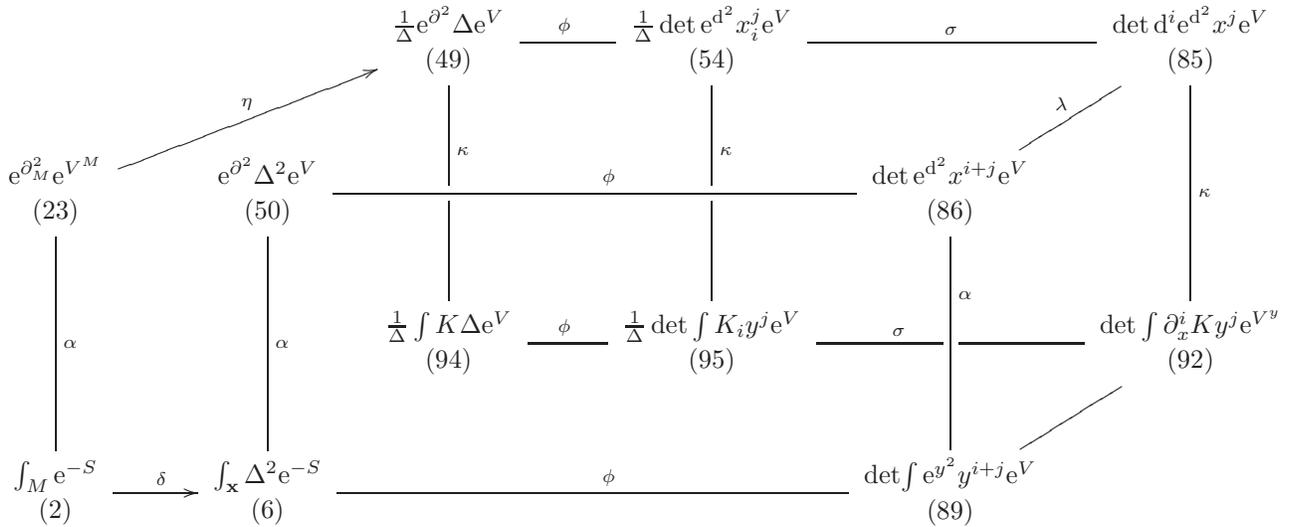
\begin{figure}[h!]
\centering
\small
\begin{adjustwidth}{-10pt}{-10pt}
\begin{align}
    \xymatrix@C-1pc{ 
    &&& 
	{\begin{array}{c} \frac1{\vd}\e^{\pd^2}\vd\e^{V} \\ \eqref{eq:diff-diagonal-new-herm} \end{array}}
    \ar@{-}[rr]^(.44){\phi}\ar@{-}[dd]|\hole^(.36){\kappa} && 
	{\diagblock{\frac{1}{\vd}\det \e^{\d^2} x_i^{j}\e^{V}}{eq:Slater-first-expr}}  \ar@{-}[rr]^{\sigma} 
	\ar@{-}[dd]|\hole^(.36){\kappa} &&
	{\diagblock{\det\d^{i} \e^{\d^{2}} x^{j} \e^{V}}{eq:Slater-diff-one-mat}} \ar@{-}[dd]^{\kappa}
	\\
	{\diagblock{\e^{\pd_M^2} \e^{V^M}}{eq:diff-one-mat-2}} \ar@{-}[dd]^{\alpha} \ar@{->}[rrru]^{\eta} &&
	{\diagblock{\e^{\pd^2}\vd^2\e^{V}}{eq:diff-diagonal-usual-herm}} \ar@{-}[rrrr]^{\phi}\ar@{-}[dd]^{\alpha}  &&&& 
	{\diagblock{\det\e^{\d^2}x^{i+j}\e^{V}}{eq:Slater-diff-one-mat-bis}}\ar@{-}[ur] ^{\lambda} \ar@{-}[dd]^(.34){\alpha}
	\\
    &&& 
    {\diagblock{\frac1{\vd} \int K \vd \e^V}{eq:diff-diagonal-new-herm-heat}} \ar@{-}[rr]^(.44){\phi} && {\diagblock{\frac1{\vd}\det\int K_i y^j \e^V} {eq:Slater-first-expr-heat}} \ar@{-}[rr]|\hole^(.39){\sigma}
    && {\diagblock{\det\int \partial^i_x K y^{j} \e^{V^y}}{eq:C-second-onematrix}}
    \\
	{\diagblock{\int_M\e^{-S}}{eq:one-mat}} \ar@{->}[rr]^{\delta} &&
	{\diagblock{\int_\x\vd^2\e^{-S}}{eq:diagonal-hermitian}} 
	\ar@{-}[rrrr]^{\phi} &&&& 
	{\diagblock{\det\!\int  \e^{y^2} y^{i+j}\e^V}{eq:slater-1-mat-usual}}  \ar@{-}[ur]
	} \nonumber
\end{align}
\end{adjustwidth}
\normalsize
\caption{Diagrammatic overview of the various expressions for one-matrix models.}
\label{fig:diagram-one-matrix}
\end{figure}

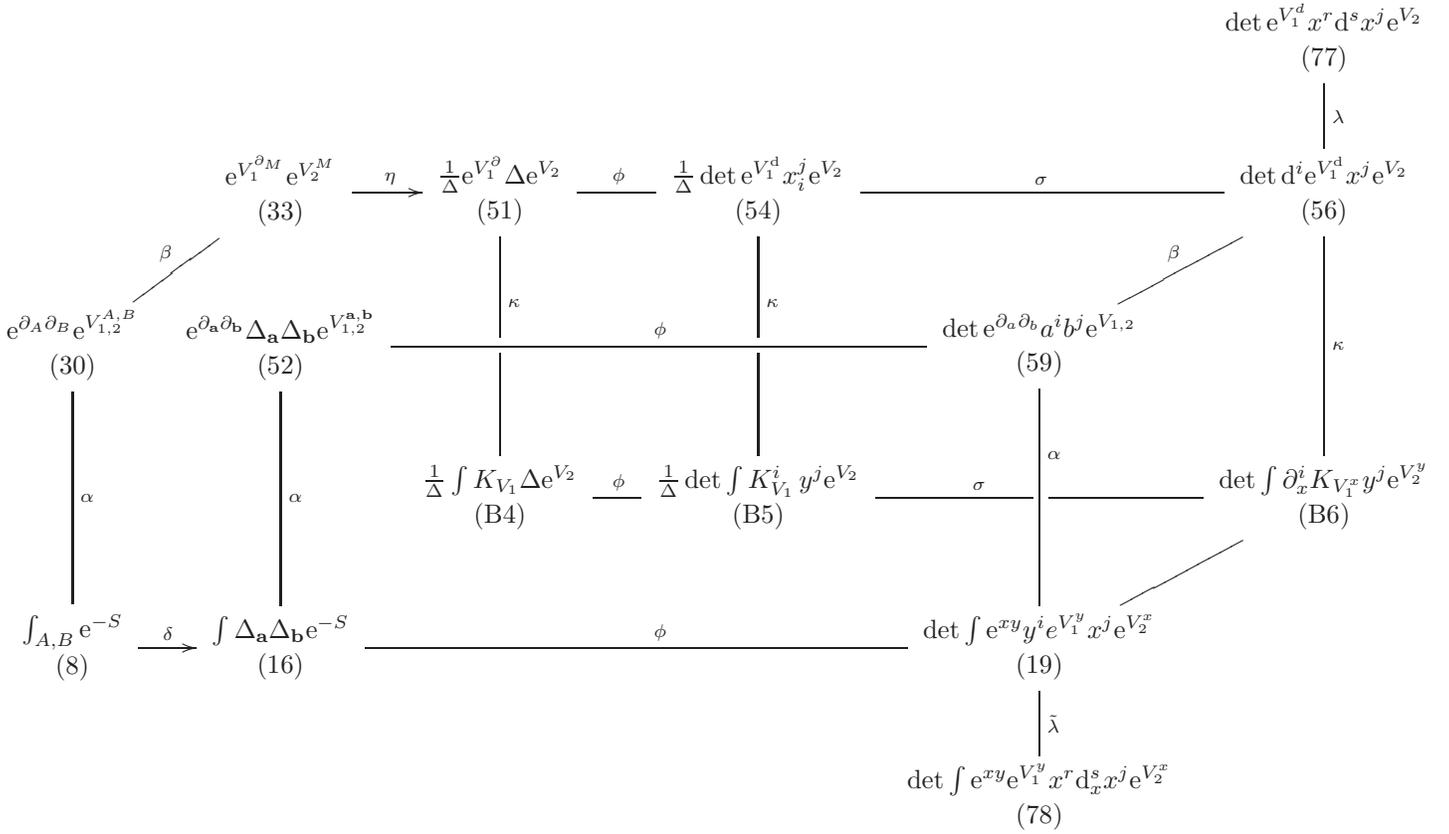
\begin{figure}[!h]
\centering
\small
\begin{adjustwidth}{-60pt}{-60pt}
\begin{align}
    \xymatrix@C-1.5pc{
    &&&&&& {\diagblock{\det \e^{V_1^d} x^r \d^s x^j \e^{V_2}}{eq:general-new-expression-p-s-diff}} \ar@{-}[d]^{\lambda}
    \\
	& {\diagblock{\e^{V_1^{\partial_M}} \e^{V_2^M}}{eq:diff-1-2-model}} \ar@{->}[r]^{\eta}
	& {\diagblock{\frac{1}{\vd} \e^{V_1^\partial} \vd \e^{V_2}}{eq:diff-diagonal-new}}
	\ar@{-}[rr]^(.46){\phi}\ar@{-}[dd]|\hole^(.36){\kappa} 
	&& {\diagblock{\frac{1}{\vd} \det \e^{V_1^\d} x_i^{j} \e^{V_2}}{eq:Slater-first-expr}} \ar@{-}[rr]^{\sigma} \ar@{-}[dd]|\hole^(.36){\kappa} 
	&& {\diagblock{\det \d^i \e^{V_1^\d} x^j \e^{V_2}}{eq:Slater-diff}}  \ar@{-}[dd]^{\kappa}
	\\
	{\diagblock{\e^{\partial_A \partial_B} \e^{V_{1,2}^{A,B}}}{eq:Two-mat-diff-form}} \ar@{-}[dd]^{\alpha}\ar@{-}[ur]^{\beta}
	& {\diagblock{\e^{\partial_\ba \partial_\bb} \vd_\ba \vd_\bb \e^{V_{1,2}^{\ba, \bb}}}{eq:diff-of-diag-two-mat}} \ar@{-}[rrrr]^{\phi}\ar@{-}[dd]^{\alpha} 
	&&&& {\diagblock{\det \e^{\partial_a \partial_b} a^i b^j \e^{V_{1,2}} }{eq:diff-of-diag-two-mat-dev}}\ar@{-}[dd]^(.36){\alpha} \ar@{-}[ur]^{\beta}
	\\
	&& {\diagblock{\frac{1}{\vd} \int K_{V_1} \vd \e^{V_2}}{eq:app-c1}} \ar@{-}[rr]^(.46){\phi} 
	&& {\diagblock{\frac{1}{\vd} \det \int K_{V_1}^i\,  y^j \e^{V_2}}{eq:app-c2}}	\ar@{-}[rr]^(.39){\sigma}|\hole 
	&& {\diagblock{\det \int \partial_x^i K_{V_1^x} y^j \e^{V_2^y}}{eq:app-c3}}
	\\
	{\diagblock{\int_{A,B}\e^{-S}}{eq:2-matrix_model}} \ar@{->}[r]^(.46){\delta}
	& {\diagblock{\int\vd_\ba \vd_\bb \e^{-S}}{eq:diag:2-mat}} \ar@{-}[rrrr]^{\phi} 
	&&&& 
	{\diagblock{\det \int \e^{xy} y^i e^{V_1^y} x^j \e^{V_{2}^{x}}}{eq:slater-2-mat-usual}} \ar@{-}[ur] \ar@{-}[d]^{\tilde \lambda}
	\\
	&&&&& {\diagblock{\det \int \e^{xy}\e^{V_1^y}x^r \d^s_x x^j \e^{V_2^x}}{eq:general-new-expression-p-s-int}}
	} \nonumber
\end{align}
\end{adjustwidth}
\normalsize
\caption{Diagrammatic overview of the various expressions for two-matrix models.}
\label{fig:diagram-two-matrix}
\end{figure}

\renewcommand{\eqref}[1]{Eq.\,(\ref{#1})}

\section{Brief presentation of the models}

We start with a reminder on the two models presenting all relevant details, in particular their perturbative expansions and diagonalizations as well as the reformulation of two-matrix models in terms of biorthogonal polynomials.

\subsection{One-matrix models}
\label{sub:oneMatDef}

\subsubsection{The model} 

The partition function of the Hermitian one-matrix model with potential $V$ is
\be
\label{eq:one-mat}
Z_V = \int_{\bH_N} \frac{\d M}{a_N} \e^{-N \tr (\frac 1 2 M^2- V(M))},
\ee
where $\mathbb{H}_N$ is the set of Hermitian matrices of size $N\times N$, the
potential is $V(x)=\sum_{k\ge 1} \frac {\lambda_k} k x^k$, the measure is $\d M=\prod_{i=1}^N \d M_{ii} \prod_{i<j} \d \Re M_{ij}\d \Im M_{ij}$, and the normalization  $a_N= {2^{-N/2}} (\pi/N)^{-{N^2}/2}$ is such that $Z_{V=0}=1$ (by explicit computation of the  Gaussian integral).
The free energy of the model is $F_V= \log Z_V$.

\subsubsection{Diagrammatic expansion}  
\label{subsub:FeynmanOne}

By expanding the exponential of the potential $V$, exchanging the summation and the integration, and using Wick's theorem, one obtains a Feynman expansion of the one-matrix model over ribbon graphs, which can be seen as graphs with an additional cyclic ordering of the edges around each vertex. Given a ribbon graph, a cyclic sequence of edges such that every edge in the sequence also precedes the edge that follows in the ordering around a vertex of the graph is called a \emph{face} (for more details see \cite{ZvonkinMapEnumeration, BouttierMapEnumeration, Eynard:2015, Eynard:CountingSurfaces, Thurigen:2021vy}). In the expansion of the partition function, the vertices of valency $k$ (that is, with $k$ edges attached) are counted with a weight $N \frac{\lambda_k} k$ and the edges with a weight $1/N$, as they correspond to propagators
\be
\label{eq:propa-one-usual}
\int_{\bH_N} \frac{\d M}{a_N} \e^{-\frac N 2 \tr ( M^2)} M_{ij} M_{kl} = \frac 1 N \delta _{i,l}\delta_{j,k}.
\ee
To each face corresponds a trace of these Kronecker symbols, resulting in an additional factor $N$ per face. The Feynman expansion of the free energy involves only connected graphs, which is not true for the partition function. 
Denoting by $V_k(\Gamma)$ the number of vertices of valency $k$ of a ribbon graph $\Gamma$, the weight (or amplitude) of a graph in the perturbative expansion is therefore 
\[
\amp(\Gamma) = \frac{N^{\chi(\Gamma)}}{|\mathrm{Aut}(\Gamma)|} \prod_{k\ge 1} \lambda_k^{V_k(\Gamma)},
\]
where $\chi(\Gamma)$ is the Euler characteristic
\[
\chi = V - E + F = 2K - 2g,
\]
in which $V=\sum_k V_k$ is the total number of vertices, $E$ is the total number of edges, $F$ the total number of faces, and $K$ the number of connected components of the ribbon graph ($K=1$ for the graph expansion of the free energy), and $g$, the genus of the ribbon graph, is a non-negative integer. The combinatorial factor $|\mathrm{Aut}(\Gamma)|$ is the number of automorphisms of the ribbon graph $\Gamma$ (see \cite{ZvonkinMapEnumeration, BouttierMapEnumeration, Eynard:2015}). 

\subsubsection{Clarification regarding matrix integrals}
\label{subsub:formal}

All matrix integrals in the paper are understood as {\it formal} integrals: They are perturbative expansions labelled by Feynman diagrams, that is, formal generating functions for ribbon graphs (see \cite{Eynard:CountingSurfaces, Orantin, Eynard:2015} and references therein). 
Every equality between partition functions is understood as equality between their perturbative expansions, term by term, regardless of the analytic properties of these series. The questions of whether the matrix integrals are well-defined, whether the series are summable when the coupling constants are in some domain and in some certain limits (e.g.~large $N$), and whether an equality between these different quantities holds non-perturbatively are not addressed in this paper.

\subsubsection{Diagonalization} 

One of the standard ways of solving this model is to first  diagonalize the Hermitian matrices as $M = U^{\dagger} X U$, where $U\in \U(N)$ and $X=\textrm{diag}(\x)$, $\x=(x_1, \ldots, x_N)\in \bR^N$. Due to the unitary invariance, the integration over unitary matrices factors out the volume of the unitary group, and the partition function is expressed as an integral over the $N$ real eigenvalues,
\be
\label{eq:diagonal-hermitian}
Z_V =  \int_{\bR^N} \frac{\d\x}{b_N}\, \vd^{2}(\x)\, \e^{-N(\frac 1 2 \x^2 - V(\x))}, 
\ee
where the measure is $\d\x=\prod_{i=1}^{N} \d x_{i}$, the normalization is $b_N={(2\pi)^{N/2}\prod_{j=1}^{N}j!}\,{N^{-N^2/2}}$ (this can be computed from Mehta's integral Eq.~ (3.3.10) in \cite{Mehta:2004wq}), 
$\Delta(\x)=\prod_{i<j} (x_j - x_i)$ is the Vandermonde determinant,  $\x^2=\sum_{i=1}^{N}x_{i}^{2}$, and we use the notation
\[V(\x)\equiv\sum_{i=1}^N V(x_i).
\]
To solve the model one may apply for example the saddle point method, loop or Schwinger-Dyson equations,  or
the method of orthogonal polynomials \cite{Eynard:2015}.
Here we focus on the steps leading to the resolution by orthogonal polynomials relying on the diagonalization.

\subsection{Two-matrix models}

\subsubsection{Partition function}  
The models we are interested in have the following form:
\be
\label{eq:2-matrix_model}
Z_{V_1, V_2} = \int_{\bH_N\times \bH_N}\frac{\d A\, \d B}{c_N} \e^{-N \tr(AB- V_1(A) - V_2(B))}
\ee
where $V_1(x)=\sum_{k\ge 1} \frac{\alpha_k} k x^k$ and $V_2(x)=\sum_{k\ge 1} \frac{\beta_k} k  x^k$, and the normalization $c_N$ is chosen such that for $V_1=V_2=0$, $Z_{0,0}=1$. {\it The Gaussian part is formal} and should be understood as follows. 
For two real variables, we formally have for $\alpha\in \mathbb{R}$:
\be
\label{eq:Gaussian-mixt-0}
\int_{\bR^2}\d x\, \d y\, \e^{- \alpha  N  xy } = \frac{2\pi} {\imath \alpha N},
\ee
and more generally for $n,m\ge 0$:
\be 
\label{eq:Gaussian-mixt-00}
\int_{\bR^2}\d x\, \d y\, \e^{- \alpha N  xy }\, x^n y^m = \frac {\delta_{n,m}  }{(- N)^n} \frac{\partial^{n}}{\partial \alpha^{n}} \int_{\bR^2}\d x\, \d y\, \e^{-  \alpha N  xy } =  \frac {\delta_{n,m}  } {(- N)^n} \frac{\partial^{n}}{\partial \alpha^{n}} \frac{2\pi}{\imath \alpha N} 
,
\ee
so that
\be
\label{eq:Gaussian-mixt}
\int_{\bR^2}\d x\, \d y\, \e^{- \c N  xy }\, x^n y^m = \delta_{n,m}\,  \frac{2\pi} \imath \,  \left(\frac 1 {\c N}\right)^{n+1}n!.
\ee
This leads to the normalization:
\be
\label{eq:normal-two-mat-mod}
c_N= \int_{\bH_N\times \bH_N}\d A\, \d B\, \e^{-N \tr(AB)}= 2^N  \left( \frac \pi {\imath N}\right)^{N^2},
\ee
and to the propagator
\be
\label{eq:propagator-two-mat-usual}
\int_{\bH_N\times \bH_N} \frac{\d A\, \d B}{c_N}\,  \e^{-N \tr(AB)}\, A_{ab}B_{cd} = \frac 1 N \delta_{a,d}\delta_{b,c}.
\ee

The integrals in \eqref{eq:Gaussian-mixt-0}, \eqref{eq:Gaussian-mixt}, \eqref{eq:normal-two-mat-mod} and \eqref{eq:propagator-two-mat-usual} are not well-defined. On the other hand, the values we set formally here are the ones needed in order to recover the correct combinatorial expansions with the correct overall normalizations of the partition functions.\footnote{The formal propagator \eqref{eq:Gaussian-mixt} can be found in [\cite{ColoringRandomTriang}, Eq.~(4.2)], but we have added the factor ${2\pi} /{\i}$ to agree with the value of \eqref{eq:2-matrix_model} when $V_1(x)=V_2(x)= - \epsilon x^2/2$, in the limit where $\epsilon \rightarrow 0$ see e.g.~\cite{Eynard:2008ht}, App. A.} In the differential formulation, the equivalent expressions are well-defined and automatically normalized, as will be detailed in the rest of the text.

\subsubsection{Diagrammatic expansion}  

The Feynman graph expansion of a two-matrix model involves a summation over ribbon graphs that have two kinds of vertices, respectively associated to the traces of the matrices $A$ and $B$, so that the edges only link vertices of different kinds (the ribbon graphs are {\it bipartite}
\footnote{As a convention, if there are quadratic terms in $V_1$ and $V_2$, we interpret them as interactions corresponding to bi-valent vertices in the graphs.}). 
Denoting by $V^A_k(\Gamma)$ and $V^B_k(\Gamma)$ the number of vertices of valency $k$ of each type of vertices of a ribbon graph $\Gamma$, the weight of a Feynman graph is given by 
\be
\label{eq:amp-two-mat-graph}
\amp(\Gamma) = \frac{N^{\chi(\Gamma)}}{|\mathrm{Aut}(\Gamma)|} \prod_{k\ge 1} \alpha_k^{V^A_k(\Gamma)}\beta_k^{V^B_k(\Gamma)},
\ee
the graphs contributing to the free-energy being connected.

\subsubsection{Diagonalization} 

The first step in the resolution is to  diagonalize $A$ and then use the Harish-Chandra--Itzykson--Zuber formula \cite{HarishChandra, Itzykson:1980hi} to diagonalize $B$:
\be 
\int_{\U(N)} \d U\,  \e^{\c\tr(AUBU^\dagger)} 
= {\c^{-\frac{N(N-1)}2}}\prod_{k=1}^{N-1} k!\  \frac{\det \{\e^{\c\, a_i b_j}\}_{1\le i,j\le N}}{\Delta(\ba)\Delta(\bb)},
\label{eq:itz-zub}
\ee
where $\d U$ is the normalized Haar measure on the unitary group $\U(N)$ and $\c$ is a complex coefficient.
As shown in \cite{Mehta:1981jm}, for $\c=N$ this leads to 
\be
\label{eq:diag:2-mat}
Z_{V_1,V_2}=  \int_{\bR^{2N}} \frac{\d\ba\,\d\bb}{N!\,d_N} \, \Delta(\ba)\Delta(\bb) \, 
\e^{-N (\ba\cdot \bb -V_1(\ba) - V_2(\bb))},
\ee
where $d_N=(2\pi / \imath)^N \prod_{j=1}^{N-1} j!\,   N^{-N(N+1)/2}$.

\subsubsection{Determinant form} 
\label{subsub:det-form-usual}

Let us detail the steps leading to the usual resolution using biorthogonal polynomials \cite{Mehta:1981jm}. The Leibniz determinant formula  
\[\label{eq:Leibnizformula}
\det(\{f_{i,j}\}_{1\le i,j \le N}) = \sum_{\sigma\in S_N} (-1)^\sigma \prod_{i=1}^N\,f_{i,\sigma(i)}
\]
where $(-1)^\sigma$ is the sign of the permutation $\sigma$,  is then used to develop the Vandermonde determinants $\vd(\x)=\det \{x_i^{j-1}\}_{1\le i,j \le N}$, so that \eqref{eq:diag:2-mat} reads:
\begin{align}
Z_{V_1,V_2}
&= \frac {1} {N!\,d_N} \sum_{\sigma, \sigma'\in S_N }(-1)^{{\sigma+ \sigma'}} \prod_{i=1}^N\int_{\bR} \d x \int_{\bR} \d y\, x^{\sigma(i)-1}y^{\sigma'(i)-1} \, \e^{-N (xy-V_1(x) - V_2(y))}\nonumber\\
&=\frac {1}{d_N}  \sum_{\sigma \in S_N }(-1)^{{\sigma}} \prod_{i=1}^N\int \d x\, \d y\,  x^{\sigma(i)-1}y^{i-1} \, \e^{-N (xy-V_1(x) - V_2(y))},
\end{align}
leading to the following expression of the partition function as a Hankel determinant (again by Eq.~\ref{eq:Leibnizformula}):
\begin{equation}
\label{eq:slater-2-mat-usual}
Z_{V_1,V_2}= \frac {1}{d_N} \det\left\{ \int \d x\, \d y\, \e^{-N xy} x^{i}  \e^{N V_1( x)} y^{j} \e^{N V_2(y)}  \right\}_{0\le i,j\le N-1}.
\end{equation}

\subsubsection{Biorthogonal polynomials}
\label{subsub:ortho-pol-usual}

Using the properties of the determinant, one can subtract from each row or column a linear combination of rows or columns of lower indices without affecting the result, so that one can replace $x^i$ and $y^{j} $ by any polynomials $P_i(x)$, $Q_j(y)$ of degrees $i$ and $j$ respectively with unit leading coefficients (such polynomials are said to be {\it monic}):
\begin{equation}
\label{eq:bil-form-int}
Z_{V_1,V_2}= \frac {1}{d_N} \det\left(\left\{ \langle P_i \vert Q_j\rangle  \right\}_{0\le i,j\le N-1}\right),  \qquad \langle P_i \vert Q_j\rangle = \int \d x\, \d y\, \e^{-N xy} P_i(x)  \e^{N V_1( x)} Q_j(y) \e^{N V_2(y)}.
\end{equation}
 In particular, the monic polynomials $P_i, Q_j$ can be chosen to be orthogonal for the 
formal symmetric bilinear form on the right-hand side of \eqref{eq:bil-form-int}, in which case the determinant \eqref{eq:slater-2-mat-usual} is the product of the diagonal terms: 
\[
\langle P_i \vert Q_j\rangle=h_i \delta_{i,j}, \qquad\qquad  Z_{V_1,V_2}= \frac {1}{d_N}\prod_{j=1}^N h_i. 
\]
Two sequences of polynomials satisfying this orthogonality relation are said to be biorthogonal. They are determined recursively for specific choices of the potentials $V_1$ and $V_2$ \cite{Mehta:1981jm, 1997NuPhB.506..633E, ColoringRandomTriang, Bouttier_2002, Bertola:2007is}.

\section{Differential reformulation of matrix integrals}
\label{sec:diff-reformulation}

In this section we show how to  reformulate two-matrix models in terms of a single matrix and its derivatives. We first review the usual differential formulations of one- and two-matrix models. Then we present the new one-matrix differential formulation of the latter. We finally show how to diagonalize the models in this formulation.

\subsection{The differential formulation}
\label{sub:diff-formulation}

It is known (see e.g.~\cite{GurauKraj}) that a Hermitian one-matrix model admits the following reformulation:
\be
\label{eq:diff-one-mat}
Z_V = \int_{\bH_N}\frac{\d M}{a_N} \e^{-N \tr (\frac 1 2 M^{2}-V(M))} = \left[ \e^{+\frac1 {2N} \tr (\frac{\pd}{\pd M})^2} \e^{N \tr\, V(M)} \right]_{M=0},
\ee
where $\tr \bigl(\frac{\pd}{\pd M}\bigr)^2=\sum_{a,b=1}^N \frac{\pd}{\pd M_{ab}} \frac{\pd}{\pd M_{ba}}$. The role of the matrices and derivatives can be exchanged:
\be
\label{eq:diff-one-mat-2}
Z_V =  \left[ \e^{N  \tr V (\frac{\pd}{\pd M})} \e^{\frac 1 {2N} \tr\, (M^2)} \right]_{M=0},
\ee
which is the formulation analogous to \eqref{eq:central-identity} for a one-matrix model.
Note that the differential expressions are clearly well-normalized, as for $V=0$, the only contributing term in the series-expansion of the exponential is 1. 

As mentioned in Sec.~\ref{subsub:formal}, the sign ``$=$'' means that the equality holds at the perturbative level: the Feynman expansions coincide term by term. While the perturbative expansion of the matrix model is obtained by expanding the exponential of $V$ in series and exchanging  the summation and the integration (Sec.~\ref{subsub:FeynmanOne}), in the differential formulation, the perturbative expansion is obtained by exchanging the summation and the evaluation $M=0$.\footnote{In the case where both sides of \eqref{eq:diff-one-mat} are well-defined non-perturbatively, one may wonder whether the two coincide.}

More precisely, recalling that $V(x)=\sum_{k\ge 1} \frac {\lambda_k} k x^k$, the perturbative expansion of the right-hand side of \eqref{eq:diff-one-mat} reads:
\be
\label{eq:dvt-proof}
\begin{split}
\left[ \e^{\frac1 {2N} \tr (\frac{\pd}{\pd M})^2} \e^{N \tr\, V(M)} \right]_{M=0} 
&=\sum_{\{n_k\ge 0\}}\prod_k \frac{(N\lambda_k/k)^{n_k}}{n_k!} \left[ \e^{\frac1 {2N} \tr (\frac{\pd}{\pd M})^2} \prod_k  \tr(M^k)^{n_k} \right]_{M=0}  \\
&= \sum_{\{n_k\ge 0\}}\prod_k \frac{(N\lambda_k/k)^{n_k}}{n_k!} \sum_{i\ge 0}\frac 1 {i!}\left[ \left(\frac1 {2N} \tr (\frac{\pd}{\pd M})^2\right)^i \prod_k  \tr(M^k)^{n_k} \right]_{M=0}  \\
&= \sum_{\{n_k\}} \prod_k \frac{(N\lambda_k/k)^{n_k}}{n_k!}\frac 1 {({\ell(\bn)}/2)!} \Biggl[\left(\frac1 {2N} \tr (\frac{\pd}{\pd M})^2\right)^{ \frac {\ell(\bn)}2} \prod_k\tr(M^k)^{n_k}\Biggr],
\end{split}
\ee
where $\ell(\bn) = \sum_k k n_k$. Developing the term originating from the propagator:
\[
\left(\tr\frac{\pd}{\pd M}\frac{\pd}{\pd M}\right)^{\frac {\ell(\bn)}2}= \sum_{p=1}^{ {\ell(\bn)}/2}\sum_{i_p, j_p=1}^N \prod_{p=1}^{ {\ell(\bn)}/2} \frac{\pd}{\pd M_{i_p j_p}}\frac{\pd}{\pd M_{j_p i_p}}, 
\]
we see that the expression between brackets in the last line of \eqref{eq:dvt-proof} is a sum over all possible ways to pair the derivatives $\frac{\pd}{\pd M_{ij}}$  and the matrix elements $M_{kl}$ originating from the interaction potential, with:
\[
\frac 1 {2N} \tr\frac{\pd}{\pd M_{i_p j_p}}\frac{\pd}{\pd M_{j_p i_p}}M_{ab}M_{cd} = \frac 1 {2N}(\delta_{i_p,a}\delta_{j_p,b}\delta_{j_p,c}\delta_{i_p,d} + \delta_{i_p,c}\delta_{j_p,d}\delta_{j_p,a}\delta_{i_p,b}).
\]
Since each $i_p$ and $j_p$ appear only in one term of the sort, the sums over $i_p,j_p$ can be carried out with the following  simplifications:
\[
\sum_{i_p,j_p} \frac 1 {2N} \tr\frac{\pd}{\pd M_{i_p j_p}}\frac{\pd}{\pd M_{j_p i_p}}M_{ab}M_{cd} = \frac 1 {N}\delta_{a,d}\delta_{b,c},
\]
where we recognize the propagator \eqref{eq:propa-one-usual}. The combinatorial factor $\frac 1 {(  {\ell(\bn)}/2)!}$ compensates for the number of permutations of the traces of pairs of derivatives (the index $p$), as all permutations give the same result. In other words, the term in square brackets in the last line of \eqref{eq:dvt-proof} coincides with a Gaussian expectation through Wick's theorem:
\[
\label{eq:Wick-One}
\frac 1 {({\ell(\bn)}/2)!} \left(\frac1 {2N} \tr\frac{\pd}{\pd M}\frac{\pd}{\pd M}\right)^{ {\ell(\bn)}/2} \prod_k\tr(M^k)^{n_k} = \int_{\bH_N} \frac{\d M}{a_N} \e^{-\frac N 2 \tr(M^2)} \prod_k\tr(M^k)^{n_k},
\]
which is expressed as a sum over ribbon graphs with $n_k$ vertices of valency $k$ and ${\ell(\bn)}/2 = \frac1 2{\sum_k k n_k}$ edges. We thus recover the usual perturbative expansion of \eqref{eq:one-mat}, which proves \eqref{eq:diff-one-mat} at the perturbative level. \eqref{eq:diff-one-mat-2} is shown the same way, with:
\[
\frac 1 {({\ell(\bn)}/2)!} \prod_k\left(\tr \left(\frac{\pd}{\pd M}\right)^k\right)^{n_k }\left(\frac1 {2N} \tr (M^2 )\right)^{ \frac{\ell(\bn)}2}  = \int_{\bH_N} \frac{\d M}{a_N} \e^{-\frac N 2 \tr(M^2)} \prod_k\tr(M^k)^{n_k}.
\]

\subsection{Differential formulation of two-matrix models}
\label{sub:diff-two-matrix}

For a two-matrix model the standard differential formulation is simply
\be
\label{eq:Two-mat-diff-form}
Z_{V_1,V_2}= \int_{\bH_N^2}\frac{\d A\, \d B}{c_N}\, \e^{-N \tr(AB-V_1(A) - V_2(B))} 
= \left[ \e^{\frac 1 N \tr\, (\frac{\pd}{\pd A} \frac{\pd}{\pd B})} \e^{N \tr\left(V_1(A)+V_2(B)\right)} \right]_{A,B=0}.
\ee
Again, the differential expression is clearly well-normalized. \eqref{eq:Two-mat-diff-form} is shown perturbatively using the fact that now the propagator  corresponds to:
\[
\sum_{i_p,j_p} \frac 1 {N} \tr\frac{\pd}{\pd A_{i_p j_p}}\frac{\pd}{\pd B_{j_p i_p}}A_{ab}B_{cd} = \frac 1 {N}\delta_{a,d}\delta_{b,c},
\]
so that through Wick's theorem, denoting by $\ell(\ba + \bb) = \sum_k k (a_k+b_k)$, we have
\begin{equation}
\label{eq:Wick-Two}
\frac 1 {\frac{\ell(\ba + \bb)} 2!}\left(\frac1 {N} \tr\frac{\pd}{\pd A}\frac{\pd}{\pd B}\right)^{\frac{\ell(\ba + \bb)} 2} \prod_k\tr(A^k)^{a_k}\tr(B^k)^{b_k} = \int \frac{\d A\d B}{c_N} \e^{ - N  \tr(AB)} \prod_k\tr(A^k)^{a_k}\tr(B^k)^{b_k},
\end{equation}
the other steps being the same as for one-matrix models. This last quantity is expressed as a sum over bipartite ribbon graphs with $a_k$ and $b_k$ vertices of valency $k$ associated to the matrix $A$ respectively $B$ and $\sum_k k a_k= \sum_k k b_k$ edges.

While the right-hand side of \eqref{eq:Wick-Two} is formal and understood {\it via} Wick's theorem and the formal propagator \eqref{eq:propagator-two-mat-usual}, the left-hand side is well-defined.

\

In the differential formulation however, it is no longer necessary to use two matrices: this is important in the integral formulation to impose 
bipartiteness of the Feynman ribbon graphs. This  bipartiteness can also be implemented in a differential formulation using the fact that derivatives only act on matrices. The resulting differential formulation thereby only involves a {\it single matrix}: 
\be
\label{eq:diff-1-2-model}
\boxed{Z_{V_1,V_2} = \left[ \e^{N \tr\, V_1(\frac 1 {\sqrt N} \frac{\pd}{\pd M})} \e^{N \tr\, V_2(\frac 1 {\sqrt{N}} M)} \right]_{M=0}.}
\ee
This can be proven by showing that this differential formula generates the same ribbon graphs, together with the same combinatorial weights \eqref{eq:amp-two-mat-graph}. It can also be proven in the differential formulation starting from the right-hand side of \eqref{eq:Two-mat-diff-form}:
\[
Z_{V_1,V_2} = \left[ \e^{\frac 1 N \tr\frac{\pd}{\pd A}\frac{\pd}{\pd B}}\,\e^{N \tr\, V_1(A)}\, \e^{N \tr\, V_2(B)} \right]_{A=B=0}= \biggl[ \left[\e^{\frac 1 N \tr\frac{\pd}{\pd A}\frac{\pd}{\pd B}}\,\e^{N \tr\, V_1(A)}\right]_{A=0} \e^{N \tr\, V_2(B)} \biggr]_{B=0},
\]
where, recalling that $\ell(\ba)=\sum_k k a_k$:
\begin{align}
 \left[\e^{\frac 1 N \tr\frac{\pd}{\pd A}\frac{\pd}{\pd B}}\,\e^{N \tr\, V_1(A)}\right]_{A=0} &= \sum_{\{a_k\ge 0 \}}\sum_{i\ge 0 } \prod_k \frac{(N \alpha_k/k)^{a_k}}{a_k!}\frac{1}{i!} \left[ \Bigl(\frac 1 N \tr\frac{\pd}{\pd A}\frac{\pd}{\pd B}\Bigr)^{i}\prod_k (\tr A^k )^{a_k}  \right]_{A=0} \\
 &= \sum_{\{a_k\ge 0 \}} \prod_k \frac{(N \alpha_k/k)^{a_k}}{a_k!}\frac{1}{\ell(\ba)!} \Bigl(\frac 1 N \tr\frac{\pd}{\pd A}\frac{\pd}{\pd B}\Bigr)^{\ell(\ba)}\prod_k (\tr A^k )^{a_k} . \nonumber
\end{align}
Assuming that we have proven that: 
\[
\label{eq:central-point-proof}
\Bigl(\tr\frac{\pd}{\pd A}\frac{\pd}{\pd B}\Bigr)^{\ell(\ba)} \prod_k \Bigl(\tr \left(A^k\right) \Bigr)^{a_k}   = \ell(\ba)! \prod_k \left(\tr \left(\frac{\pd}{\pd B}\right)^k\right)^{a_k}, 
\]
we obtain 
\be
\label{eq:proof3C}
\left[\e^{\frac 1 N \tr\frac{\pd}{\pd A}\frac{\pd}{\pd B}}\,\e^{N \tr\, V_1(A)}\right]_{A=0}  = \e^{N V_1(\frac 1 N \frac{\pd}{\pd B})}, 
\ee
and thereby 
\[
Z_{V_1,V_2} = \left[ \e^{N \tr\, V_1(\frac 1 N \frac{\pd}{\pd B})} \e^{N \tr\, V_2(B)} \right]_{B=0}. 
\]
The sought formula \eqref{eq:diff-1-2-model} is obtained by change of variable $M=\sqrt N B$.\footnote{Note that in this differential formulation, changes of variables are made without ``Jacobian'', since both sides are clearly normalized.}

Let us now prove \eqref{eq:central-point-proof}.
We write $$\left(\tr\frac{\pd}{\pd A}\frac{\pd}{\pd B}\right)^{\ell(\ba)}= \sum_{p=1}^{\ell(\ba)}\sum_{i_p, j_p=1}^N \prod_{p=1}^{\ell(\ba)}\, \frac{\pd}{\pd A_{i_p j_p}}\frac{\pd}{\pd B_{j_p i_p}}. $$
For every value taken by $\{i_p, j_p\}_{p}$, the quantity 
$ \prod_{p=1}^{\ell(\ba)} \frac{\pd}{\pd A_{i_p j_p}}\frac{\pd}{\pd B_{j_p i_p}} \prod_k \left(\tr\left(A^k\right) \right)^{a_k}$ is the sum over all possible ways of assigning all the derivatives $\frac{\pd}{\pd B_{j_p i_p}}\frac{\pd}{\pd A_{i_p j_p}}$ to the matrix elements $A_{kl}$ in the product of traces. Every time a term of the form $\frac{\pd}{\pd B_{j_p i_p}}\frac{\pd}{\pd A_{i_p j_p}}$  acts on a matrix element $A_{kl}$, the latter is replaced in the product of traces by 
$\frac{\pd}{\pd B_{l k}}\delta_{i_p, k}\delta_{j_p, l}.$ All the indices $i_p, j_p$ belong to a single Kronecker symbol, so that the sum  $\sum_{p=1}^{\ell(\ba)}\sum_{i_p, j_p=1}^N$ simply eliminates all Kronecker symbols. We are left with 
\[
\Bigl(\tr\frac{\pd}{\pd A}\frac{\pd}{\pd B}\Bigr)^{\ell(\ba)} \prod_k \tr (A^k) ^{a_k} = \hspace{-0.2cm}\sum_{\substack{{\textrm{possible}}
\\
{\textrm{ assignments}}}} \prod_k \left(\tr \Bigl(\bigl(\frac{\pd}{\pd B} \bigr)^T\Bigr)^k \right)^{a_k} = \ell(\ba)!\, \prod_k \left(\tr \bigl(\frac{\pd}{\pd B} \bigr)^k \right)^{a_k},
\]
where the last equality holds because there are $(\sum_k k a_k)!$ ways of assigning the derivatives $\frac{\pd}{\pd B_{j_p i_p}}\frac{\pd}{\pd A_{i_p j_p}}$ to the matrix elements $A_{kl}$ in the product of traces, and the result for each assignment is independent of the specific assignment chosen. This concludes the proof.

\subsection{Diagonalization in the differential formulation}
\label{sub:diff-formulation-diagonalization}

It is known \cite{Itzykson:1980hi, Zuber:2008kn} that for Hermitian matrices $A$, 
the differential operator 
$D_k(\frac{\pd}{\pd A}) = \tr (\frac{\pd}{\pd A})^k$ 
acting on  $U(N)$-invariant functions\footnote{{One may of course wonder if the developments of the paper still hold for matrix models involving real symmetric matrices.  The diagonalization would require a  formula analogous to \eqref{eq:diagonaldiffoperator}, but for the derivatives of  a real symmetric matrix-valued $O(N)$-invariant function.  Such a formula is not known,  because the equivalent of the HCIZ formula \eqref{eq:itz-zub} with the integration over $U(N)$ matrices replaced with an integration over $O(N)$ matrices is known for \emph{skew-}symmetric matrices $A,B$, whereas the case of symmetric matrices is much more involved \cite{Zuber:2008kn}. }} 
diagonalizes as
\be
\label{eq:diagonaldiffoperator}
D_k\Bigl(\frac{\pd}{\pd a_i}\Bigr) = \frac 1 {\vd(\ba)} \sum_{i=1}^{N}\frac{\pd^{k}}{\pd a_{i}^{k}} \vd(\ba). 
\ee
This implies in particular that the differential operator in \eqref{eq:diff-1-2-model} has the diagonal form
\be 
\label{eq:diag-pot-diff}
\tr\, V_1\left( \frac1{\sqrt{N}} \frac{\pd}{\pd M} \right) 
= \sum_{k\ge1}\frac{\alpha_k}{k \sqrt{N}^k} D_k\Bigl(\frac{\pd}{\pd M}\Bigr) 
= \frac 1 {\vd(\bfm)} \sum_{i=1}^N V_1\left(\frac1{\sqrt{N}} \frac{\pd}{\pd m_i} \right) \vd(\bfm),
\ee
since it acts on $\exp(V_2( M))$, which depends on traces of $M$ and is thus $U(N)$-invariant.

Since this is only briefly mentioned in the final remarks of \cite{Zuber:2008kn}, let us make the argument more precise here.
One defines the operator $\widehat{D}_k$ for some positive integer $k$ as the operator which acts on the exponential trace of the product of two Hermitian matrices $A$ and $B$ as
\[\label{eq:defdiffoperator}
\widehat{D}_k  \e^{\c\tr A B} \equiv \c^k \tr(B^k) \e^{\c\tr A B}
\]
and $\c$ is a complex coefficient. 
From this definition it follows that this operator has the explicit representation as a functional of matrix derivatives
\be
\label{eq:def-Dk-1}
D_k\Bigl(\frac{\pd}{\pd A}\Bigr) = \tr \Bigl(\frac{\pd}{\pd A}\Bigr)^k 
= \sum_{i_1,...,i_k} \frac{\pd}{\pd A_{i_1 i_2}}\frac{\pd}{\pd A_{i_2 i_3}}\dots\frac{\pd}{\pd A_{i_k i_1}} \, 
\ee
on functions of Hermitian matrices $f(A)$ whose function space is spanned by the basis $\{\e^{\c\tr A B}\}_{B\in\mathbb{H}_N}$. For $\c=\i$, any Hermitian matrix-valued function belongs to that space by Fourier transformation on Hermitian matrices (see e.g.~\cite{Harmonic})
\be
\label{eq:Fourier-f-invariant}
f(A) =\frac {1}{2^N \pi^{N^2}} \int_{\mathbb{H}_N} \d B\, \tilde{f}(B) \,\e^{\i\tr A B}\,.
\ee

Consider now the representation of the operator $\hat{D}_k$ on the subspace of  $\U(N)$-invariant functions $f(A)=f(UAU^{-1})$, or equivalently 
\be 
\label{eq:After-Fourier-HCIZ}
f(A)=\int_{\U(N)}\d U\,f(UAU^{-1}) 
= \frac {1}{2^N \pi^{N^2}}\int_{\mathbb{H}_N} \d B \tilde{f}(B) \int_{\U(N)} \d U\,\e^{\c\tr (U A U^{-1} B)}\, .
\ee
The domain of such invariant functions is $N$-dimensional, and the functions can be represented on diagonal matrices. 
The defining equation \eqref{eq:defdiffoperator} on the invariant subspace thus has the form
\be
\widehat{D}_k \int_{\U(N)} \d U \e^{\c\tr (U A U^{-1} B)} 
\equiv \c^k \tr(B^k) \int \d U \e^{\c\tr (U A U^{-1} B)}
\]
Using the Itzykson-Zuber integral formula \eqref{eq:itz-zub},
it follows directly from a Laplace expansion of the determinant,
\begin{align}\label{eq:Laplaceexpansion}
\sum_l\Bigl(\frac{\pd}{\pd a_l}\Bigr)^k \det(\e^{\c a_m b_n})_{m,n} 
&= \sum_l \Bigl(\frac{\pd}{\pd a_l}\Bigr)^k \sum_{j}(-1)^{l+j} \e^{\c a_l b_j} \det(\e^{\c a_m b_n})_{\substack{m\ne l,\\n\ne j}} \\
&= \sum_{j} (\c b_j)^k \sum_l 
(-1)^{j+l} \e^{\c a_l b_j} \det(\e^{\c a_m b_n})_{\substack{m\ne l,\\n\ne j}}
= \c^k\tr(B^k) \det(\e^{\c a_m b_n})_{m,n} \nonumber\, ,
\end{align}
that the functional of eigenvalue derivatives $D_k(\frac{\pd}{\pd a_i})$ \eqref{eq:diagonaldiffoperator} is the representation of $\widehat{D}_k$ on $\U(N)$-invariant functions in diagonalized variables, that is on eigenvalues.

\

Another equivalent way to prove this formula is to see the function $f$  as a function of the matrix elements $\{A_{i,j}\}$ in the canonical basis and consider the eigenvalues $\{a_1,\ldots,a_N\}$ of $A\in\bH_N$,  together with $N(N-1)$ variables  that determine uniquely $U\in \U(N)$ such that $A =U \textrm{Diag}(\{a_i\})U^\dagger$,  as new variables after a change of variables (see the detailed discussion in Chapter 3 of \cite{Mehta:2004wq}). Differentiating $\sum_{l,k} U^{\dagger}_{il} A_{lk}U_{kj}$ with respect to the $a_p$, we find
$
\sum_{l,k} U^{\dagger}_{il} \frac{\partial A_{lk}}{\partial a_p}U_{kj} = \delta_{ij} \delta_{jp},
$
which is inverted as
$
 \frac{\partial A_{ij}}{\partial a_p} = U_{ip}U^{\dagger}_{pj}. 
$
From this, the derivative of any function of $A$,
$ \frac{\partial}{\partial a_p} f(\{A_{ij}\})$ is well-defined, and we may verify explicitly that both the right-hand-side $D^{(1)}$ of \eqref{eq:def-Dk-1} acting on a unitary invariant function  expressed as \eqref{eq:Fourier-f-invariant}, and the right-hand-side $D^{(2)}$ of \eqref{eq:diagonaldiffoperator} acting on \eqref{eq:After-Fourier-HCIZ} expressed in eigenvalue-variables using \eqref{eq:itz-zub} give the same expression: 
\[
D^{(1)} f(A) 
=  
D^{(2)} f (A) 
= 
\frac {\i^k}{2^N \pi^{N^2}} \int_{\bH_N} \d B \tilde f (B) \tr(B^k)\e^{\imath \tr(AB)}.
\]
This requires using the fact that the Fourier transform $\tilde f (B)$ is also unitary invariant.

\

We use this formula to diagonalize the differential formulations.
For a one-matrix model, applying \eqref{eq:diagonaldiffoperator} to \eqref{eq:diff-one-mat}, we obtain
\be
\label{eq:diff-diagonal-new-herm}
Z_V=\left[ \e^{\frac{1}{2N} \tr \frac{\pd^{2}}{\pd M^{2}}} \e^{N V(M)} \right]_{M=0} 
= \left[ \frac 1 {\vd(\x)} 
\e^{\frac 1 {2N}(\frac{\pd}{\pd \x})^2}
\vd(\x) \,\e^{N V(\x)} \right]_{\x=0}
\ee
where $(\frac{\pd}{\pd \x})^2 = \sum_i \frac{\pd^{2}}{\pd x_{i}^{2}}$.
The formula is clearly well-normalized, since $(\frac{\pd}{\pd \x})^2 \vd(\x) = 0$.
Note that the differential reformulation of the usual integral over eigenvalues \eqref{eq:diagonal-hermitian} reads
\be
\label{eq:diff-diagonal-usual-herm}
Z_V = \int_{\bR^N} \frac{\d \x}{b_N} \e^{- \frac {N} {2} \x^2} \vd^{2}(\x)\, \e^{NV(\x)} = e_N\left[ \e^{\frac 1 {2N}(\frac{\pd}{\pd \x})^2}\vd^{2}(\x)\, \e^{N V(\x)} \right]_{\x=0} ,
\ee
where the normalization $
e_N = N^{N(N-1)/2} / \prod_{j=1}^N j!
$  is computed in Appendix \ref{app:normalizations}.

\

For the diagonalization of two-matrix models it is crucial that we have a differential formulation  using only one matrix such as \eqref{eq:diff-1-2-model}.
It is not possible to simultaneously diagonalize differential operators involving two matrices such as $\tr\, (\frac{\pd}{\pd A} \frac{\pd}{\pd B})$ in \eqref{eq:Two-mat-diff-form}, at least not by the arguments used here.
However, the differential formulation in terms of a single matrix allows us to diagonalize  two-matrix models.

Applying \eqref{eq:diag-pot-diff} to \eqref{eq:diff-1-2-model} yields
\be
\label{eq:diff-diagonal-new}
\boxed{Z_{V_1,V_2}=  \left[ \frac 1 {\vd(\x)} \e^{N V_1( \frac 1 {\sqrt N} \frac{\pd}{\pd \x})} \vd(\x) \,\e^{NV_2(\frac 1 {\sqrt N}  \x)} \right]_{\x=0}}
\, .\ee
For comparison,  the differential formulation of the usual integral over eigenvalues \eqref{eq:diag:2-mat} is
\be
\label{eq:diff-of-diag-two-mat}
Z_{V_1,V_2}=  e_N\left[ \e^{\frac 1 N \sum_{i=1 }^N\frac{\pd}{\pd a_i}\frac{\pd}{\pd b_i}}\,\Delta(\ba)\Delta(\bb)\e^{N V_1(\ba)}\, \e^{N V_2(\bb)} \right]_{\ba=\bb=0}
\ee
where the normalization is the same as for the one-matrix model (see Appendix \ref{app:normalizations}).

The equivalence between \eqref{eq:diff-diagonal-new-herm} and \eqref{eq:diff-diagonal-usual-herm} on one hand, and between \eqref{eq:diff-diagonal-new}  and \eqref{eq:diff-of-diag-two-mat} on the other hand is not manifest in this form, however it becomes clear when formulated in terms of Slater determinants, as detailed later in  Sec.~\ref{subsub:Equivalence-diag-Twomat} for the two-matrix case and in Sec.~\ref{subsub:invert-diff-diag-gaussian} for the one-matrix model.

\section{Expressions as Slater determinants and applications}

\subsection{Equivalent formulations of a two-matrix model}

In this first subsection, we detail the steps leading to the Slater determinant formulation of the partition functions of two-matrix models, in differential formulation.
{The same holds for the one-matrix models,  as the special case $V_1(x)=\frac{x^2}{2}$ (cf.~Sec.~\ref{subsub:invert-diff-diag-gaussian}).}

\subsubsection{Determinant form} 
\label{subsub:slater-diff-two-mat}
Starting from  \eqref{eq:diff-diagonal-new}:
\[
Z_{V_1,V_2}=   \left[ \frac 1 {\vd(\x)} \left(\prod_{i=1}^N \e^{N V_1( \frac 1 {\sqrt N} \frac{\pd}{\pd x_i})}\right) \vd(\x) \left( \prod_{i=1}^N \e^{NV_2(\frac 1 {\sqrt N} x_i)}\right) \right]_{\x=0},
\]
 we incorporate the products on the left and right of the Vandermonde determinant into a Slater determinant: 
\be
\label{eq:Slater-first-expr}
Z_{V_1,V_2}=   \left[ \frac 1 {\vd(\x)} \det \left\{ \e^{N V_1( \frac 1 {\sqrt N} \frac{\d}{\d x_i})} x_i^{j-1} \e^{N V_2(\frac 1 {\sqrt N} x_i)} \right\}_{1\le i,j\le N}  \right]_{\x=0}.
\ee

We can extract the symmetric part of the determinant by using the following identity (\cite{Hua}, Thm.~1.2.4 p.~24):
\be
\label{eq:identity-diff-det}
\left[\frac{1}{\vd(\x)} \det_{i,j}\bigl(f_j(x_i)\bigr)\right]_{\forall i,\, x_i=x} = f_N\,\det_{i,j}\left (f_j^{(i-1)}(x)\right),
\ee
with $f_N= 1/\prod_{j=1}^{N-1} j !$.\footnote{The factor $(-1)^{\frac {N(N-1)}{2}}$ in the reference is due to a different convention in the definition of the Vandermonde determinant.} This leads to the following Slater determinant:

\begin{equation}
\label{eq:Slater-diff}
Z_{V_1,V_2}= f_N \det\left\{ \left[ \dxi{i}  \e^{N V_1(\frac 1 {\sqrt N} \dx)} x^{j} \e^{N V_2(\frac 1 {\sqrt N} x)} \right]_{x=0} \right\}_{0\le i,j \le N-1}.
\end{equation}
This is again well-normalized, since for $V_1=V_2=0$ the determinant is $\prod_{i=1}^{N-1} i!$.

\subsubsection{Biorthogonal polynomials}  

As in Sec.~\ref{subsub:ortho-pol-usual}, $\dxi{i} $ and $x^{j} $ can be replaced in \eqref{eq:Slater-diff} by any monic polynomials $P_i(\dx)$, $Q_j(x)$ respectively of degrees $i$ and $j$:
\begin{equation}
\label{eq:Slater-diff-poly}
Z_{V_1,V_2}= f_N \det\left\{ \left[ P_i\Bigl(\dx\Bigr) \e^{N V_1( \frac 1 {\sqrt N} \frac{d}{d x})} Q_j(x) \e^{N V_2(\frac 1 {\sqrt N} x)} \right]_{x=0} \right\}_{0\le i,j \le N-1},
\end{equation}
the differential formulation of the bilinear form of \eqref{eq:bil-form-int} being:
\begin{equation}
\langle f \vert g \rangle =  c_1 \left[ f \biggl(\frac 1 {\sqrt N}\dx\biggr) \e^{N V_1( \frac 1 {\sqrt N}\dx)} g\biggl(\frac x {\sqrt N}\biggr) \e^{N V_2(\frac 1 {\sqrt N} x)} \right]_{x=0}.
\end{equation}
From this, one can apply the method of biorthogonal polynomials. 

\subsubsection{Equivalence between the diagonalized formulations for two-matrix models}
\label{subsub:Equivalence-diag-Twomat}

Expanding the Vandermonde determinants in \eqref{eq:diff-of-diag-two-mat} as in Sec.~\ref{subsub:det-form-usual}, we obtain:
\be
\label{eq:diff-of-diag-two-mat-dev}
Z_{V_1,V_2}=  e_N N!  \det\left\{ \left[ \e^{\frac 1 N \frac{\partial}{\partial a} \frac{\partial}{\partial b}}a^i  \e^{N V_1(a)} b ^{j} \e^{N V_2(b)} \right]_{a=b=0}\right\}_{0\le i,j \le N-1}.
\ee
This is seen to be equivalent to \eqref{eq:Slater-diff},  since, from \eqref{eq:proof3C} which also holds when $A,B$ are real variables (and seeing $N$ as a real coefficient):
\[
\left[ \dxi{i}  \e^{N V_1(
\frac 1 {\sqrt N} 
\frac{d}{d x})} x 
^{j} \e^{N V_2(\frac 1 {\sqrt N} x)} \right]_{x=0} = N^{i/2} \left[ \e^{\frac 1 {\sqrt N} \frac{\partial}{\partial a} \frac{\partial}{\partial b}}a^i  \e^{N V_1(a)} b ^{j} \e^{N V_2(\frac 1 {\sqrt N}b)} \right]_{a=b=0} ,
\]
which in turn by the change of variable $b = \sqrt N b'$ is equal to 
$
 N^{\frac{i+j}2} [ \e^{\frac 1  N \frac{\partial}{\partial a} \frac{\partial}{\partial b}}a^i  \e^{N V_1(a)} b ^{j} \e^{N V_2(b)} ]_{a=b=0}.
$
The normalizations do agree since $f_N = e_N N! / N^{N(N-1)/2}$.

\eqref{eq:diff-of-diag-two-mat-dev}  is in turn seen to be equivalent to the Hankel determinant \eqref{eq:slater-2-mat-usual}, as from \eqref{eq:Two-mat-diff-form} for $N=1$:
\be \left[ \e^{\frac 1  N \frac{\partial}{\partial a} \frac{\partial}{\partial b}}a^i  \e^{N V_1(a)} b ^{j} \e^{N V_2(b)} \right]_{a=b=0} = \int_{\bR^2} \frac{\d x\, \d y}{c_1}\, \e^{-N xy} x^{i}  \e^{N V_1( x)} y^{j} \e^{N V_2(y)},
\ee
where we recall that $c_1=\frac{2\pi}{i N}$, so that $e_N N! / (c_1)^N = N^{\frac{N(N-1)} 2}f_N/(c_1)^N=1/d_N$.

\subsubsection{Expansion over Schur functions}
Performing steps similar to Sec.~\ref{subsub:det-form-usual} but in the differential formulation and the opposite direction, we may re-express \eqref{eq:Slater-diff} as:
\be
\label{eq:Orlov}
Z_{V_1,V_2}= \frac{f_N}{N!} \left[{\Delta\biggl(\frac{\pd}{\pd \x}\biggr)} \e^{NV_1( \frac 1 {\sqrt N}\frac{\pd}{\pd \x})} \vd(\x) \,\e^{NV_2(\frac 1 {\sqrt N} \x)} \right]_{\x=0}.
\ee
As explained in (\cite{Orlov:2002wg}, Eq~(2.2.4)), the expression \eqref{eq:Orlov} is easily seen --- by its action on Schur functions --- to be a scalar product on symmetric functions, with 
\[Z_{V_1,V_2}= f_N\left \langle \exp\left(N V_1\left( \frac 1 {\sqrt N}\,\bigcdot\,\right)\right), \exp\left(N V_2\left( \frac 1 {\sqrt N}\,\bigcdot\,\right)\right) \right\rangle . \] Using the orthogonality of Schur functions for this scalar product and the Cauchy-Littlewood formula, one obtains directly an explicit double-series expansion over Schur functions  (Eq.~(2.2.13) in \cite{Orlov:2002wg}), which guarantees the fact that $Z_{V_1,V_2}$ is a tau function of the Toda hierarchy.

\subsection{Integration over one variable in the determinant form of two-matrix models}
\label{sub:integration}

In order to verify the equivalence --- in  differential formulation --- between \eqref{eq:diff-diagonal-new-herm} and \eqref{eq:diff-diagonal-usual-herm}, it is required to go to Slater determinant form, and then transform certain derivatives to variables.
We do this for general $V_1$, as the computation is the same in differential formulation for general $V_1$ or for $V_1(x)=x^2/2$ (corresponding to the one-matrix models). 
We also detail the computation in the integral formulation for general $V_1$, which is naturally more involved than for $V_1(x)=x^2/2$, in which case it is a simple Gaussian integration. This leads to a new determinant formulation of the partition functions of two-matrix models, and we comment on its potential use for resolutions involving orthogonal polynomials instead of biorthogonal polynomials. 

\subsubsection{General computation}
\label{subsubsec:gen-comp}
We start from \eqref{eq:Slater-diff}:
\be
\label{eq:ex-p}
Z= f_N \det_{i,j}\left\{ \left[ \dxi{i}  \e^{ NV_1(\frac1 {\sqrt N}\dx)} x^{j} \e^{N V_2( \frac x{\sqrt N})} \right]_{x=0} \right\} =  f_N \det_{i,j}\left\{ \left[ \dxi{i} \e^{ N V_1(\frac 1 N \dx)} x^{j} \e^{N V_2( x)} \right]_{x=0} \right\},
\ee
where $0\le i,j\le N-1$, by change of variable.\footnote{The factors  $\sqrt N^{j-i}$ from the change of variable compensate when factorized out of the determinant} In this subsection, we assume $V_1$ to be a polynomial:
\[
V_1(x)=\sum_{k =1}^p \frac{\alpha_k} k x^k, \qquad p\ge 1.
\]
We write, as in \eqref{eq:proof3C} but for real variables:
\be
\label{eq:second-expr-p}
\left[ \dxi{i}  \e^{N V_1(\frac 1 N\dx)} x^{j} \e^{N V_2( x)} \right]_{x=0} =\left[ \left[\e^{\dx\dy}y^i  \e^{N V_1(\frac 1 N y)}\right]_{y=0} x^{j} \e^{N V_2( x)} \right]_{x=0}.
\ee
We first notice that:
\be 
\dyi{r}   \e^{N V_1(\frac 1 N y)} = (N^{1-p}\alpha_p)^r S_{(p-1)r}(y)\,   \e^{NV_1(y)},
\ee
where $S_{(p-1)r}(y)$ is a monic polynomial of degree $(p-1)r$. \\

Let us consider the row $i=(p-1)r+s$, where $s\in\{0,1,\ldots, p-1\}$. We can replace in the determinant the right-hand side of \eqref{eq:second-expr-p}  by $\Bigl[\Bigl[\e^{\dx\dy} y^s S_{(p-1)r}(y)   \e^{NV_1(\frac 1 Ny)}\Bigr]_{y=0} x^{j} \e^{N V_2( x)} \Bigr]_{x=0}$ and therefore by
\be
\label{eq:interm-gen-comp}
Z=  f_N \det\left\{ (N^{1-p}\alpha_p)^{-r(i)}\left[\left[\e^{\dx\dy}  y^{s(i)} \dyi{r(i)}   \e^{NV_1(\frac 1 Ny)}\right]_{y=0} x^{j} \e^{N V_2( x)} \right]_{x=0} \right\}_{0\le i,j \le N-1}.
\ee
 We now show that:
\be 
\label{eq:point-to-prove-p-s}
\left[ \left[\e^{\dx\dy} y^s \dyi{r}   \e^{ NV_1(\frac 1 N y)}\right]_{y=0} x^{j} \e^{N V_2( x)} \right]_{x=0} = \left[ \e^{NV_1(\frac 1 N\dx)} x^r \dxi{s} \Bigl[x^{j} \e^{N V_2( x)} \Bigr]\right]_{x=0},
\ee
that is, just as for \eqref{eq:proof3C} or  \eqref{eq:second-expr-p}, the result of the inner bracket $[...]_{y=0}$ on the left hand side is obtained by ``replacing'' the $y$'s by $\dx$'s, but also the $\dy$'s by $y$'s, while reversing the left-to-right ordering.

Developing first the right hand side, we see that:
\[
\left[ \e^{NV_1(\frac 1 N\dx)} x^r \dxi{s} \left[x^{j} \e^{N V_2( x)} \right]\right]_{x=0} = \sum_{\{n_k\ge 0\}_k} \prod_{k=1}^p \frac {(N^{1-k}\alpha_k/k)^{n_k}} {n_k!}  \left[ \dxi{\ell(\bn)} x^r \dxi{s} \left[x^{j} \e^{N V_2( x)} \right] \right]_{x=0},
\]
where $\ell(\bn) = \sum_{k=1}^p k n_k$. The bracket on the right hand side of this equation is non-vanishing only if $\ell(\bn) = \sum_{k=1}^p k n_k\ge r$ and $r$ of the derivatives act on $x^r$, in which case: 
\[
\left[\dxi{\ell(\bn)} x^r \dxi{s} \left[x^{j} \e^{N V_2( x)} \right] \right]_{x=0} = r! \binom{\ell(\bn)}{r}\left[  \dxi{\ell(\bn)-r+s} x^{j} \e^{N V_2( x)} \right]_{x=0}. 
\]
The right hand side of \eqref{eq:point-to-prove-p-s} is therefore equal to:
\be
\label{eq:expansion-proof-gen-comp}
 \sum_{\bigl\{n_k \ge 0\; \mid\; \ell(\bn) \ge\, r\bigr\}}\prod_{k=1}^p \frac {(N^{1-k}\alpha_k/k)^{n_k}} {n_k!} \frac {\ell(\bn) !}{ (\ell(\bn)-r)! }  \left[  \dxi{\ell(\bn)-r+s}x^{j} \e^{N V_2( x)} \right]_{x=0}.
\ee

We now focus on the left hand side of \eqref{eq:point-to-prove-p-s}:
\be
\label{eq:LHS-gen-comp}
\left[\e^{\dx\dy} y^s \dyi{r}  \e^{NV_1(\frac 1 N y)}\right]= \sum_{q\ge 0} \frac 1 {q!}  \left[\dyi{q} y^s \dyi{r} \e^{NV_1(\frac 1 N y)}\right] \dxi{q} 
= \sum_{q\ge s} \frac {s!} {q!}  \binom{q}{s}\hspace{-0.1cm}\left[  \dyi{q+r-s}   \e^{NV_1(\frac 1 N y)}\right] \dxi{q},
\ee
in which the three brackets are evaluated in $y=0$. We develop:
\begin{align}
\label{eq:LHS-gen-comp-21}
\left[  \dyi{q+r-s}   \e^{NV_1(\frac 1 N y)}\right]_{y=0} & = \sum_{\{n_k\ge 0\}_k} \prod_{k=1}^p \frac {(N^{1-k}\alpha_k/k)^{n_k}} {n_k!} \left[  \dyi{q+r-s}   y^{\ell(\bn)}\right]_{y=0}\\  &= \sum_{\substack{{\{n_k\ge 0\}\textrm{ s.t.}}\\[+0.5ex]{ \ell(\bn) = q + r - s}}} \prod_{k=1}^p \frac {(N^{1-k}\alpha_k/k)^{n_k}} {n_k!} \ell(\bn) !.
\label{eq:LHS-gen-comp-22}
\end{align}
Replacing this in \eqref{eq:LHS-gen-comp} and  then \eqref{eq:LHS-gen-comp} in the left hand side of \eqref{eq:point-to-prove-p-s} leads to the same expansion \eqref{eq:expansion-proof-gen-comp}. 
We have shown so far that: 
\begin{mdframed}[nobreak=true, innerbottommargin=5pt, innertopmargin=-7pt]
\[
Z_{V_1, V_2} = f_N\det\left\{ M_{i,j} \right\}_{0\le i,j \le N-1},
\]
where:
\begin{align}
\label{eq:general-new-expression-p-s-diff}
M_{(p-1)r + s\; ,\; j} &= ({N^{p-1}}/ {\alpha_p})^{r} \left[ \e^{NV_1(\frac 1 N \dx)} x^r \dxi{s} \Bigl[x^{j} \e^{N V_2( x)} \Bigr]\right]_{x=0}  \\[+1ex]&= ({N^{p-1}}/ {\alpha_p})^{r} \int_{\bR^2}\frac{\d x\, \d y}{2\pi / \imath}\, \e^{-xy} \e^{NV_1(\frac 1 N y)} x^r \dxi{s} \Bigl[x^{j} \e^{N V_2( x)} \Bigr].
\label{eq:general-new-expression-p-s-int}
\end{align}
\end{mdframed}

The equivalence between \eqref{eq:general-new-expression-p-s-diff} and \eqref{eq:general-new-expression-p-s-int} is the usual one-matrix differential formulation of two-matrix models of Sec.~\ref{sub:diff-two-matrix} but for $A,B$ two real variables, and seeing $x^r \dxi{s} [x^{j} \e^{N V_2( x)}]$ as a function of $x$.

\subsubsection{Computation in the integral formulation}
 Starting from \eqref{eq:slater-2-mat-usual}, we change variable as $y' = N y $ and include the normalization by $2\pi/\imath$:
 \begin{equation}
Z_{V_1,V_2}= f_N \det\left\{ \int_{\bR^2} \frac{\d x\, \d y}{2\pi / \imath}\, \e^{- xy} x^{i}  \e^{N V_1( \frac 1 N x)} y^{j} \e^{N V_2(y)}  \right\}_{0\le i,j\le N-1},
\end{equation}
where we have used the fact that $f_N = (2\pi/\imath)^N d_N^{-1} N^{-\frac{N(N+1)}2}$.
The steps up to equation \eqref{eq:interm-gen-comp} are precisely the same in integral form, so we do not detail them again. While a perturbative proof of \eqref{eq:point-to-prove-p-s} is certainly possible, it does not seem to us that the proof above can directly be translated in integral form  (in particular the steps \eqref{eq:LHS-gen-comp} to \eqref{eq:LHS-gen-comp-22}), so we propose a different proof. In integral form,  \eqref{eq:interm-gen-comp} reads
\be 
Z_{V_1,V_2}= f_N \det \left\{(N^{1-p}\alpha_p)^{-r(i)} \int_{\bR^2} \frac{\d x\, \d y}{2\pi / \imath}\, \e^{- xy} y^{s(i)} \left[ \dyi{r(i)}   \e^{ NV_1(y)}\right] x^{j} \e^{N V_2( x)} \right\}_{0\le i,j\le N-1}.
\ee
By performing $r$ integrations by part:
\be 
\label{eq:point-proof-int-form}
\int \d x\, \d y\, \e^{- xy} y^s \left[\dyi{r}   \e^{ NV_1(y)}\right] x^{j} \e^{N V_2( x)}  = (-1)^r \int \d x\, \d y\, \left[\dyi{r} \e^{-xy} y^s\right] \e^{ NV_1(y)}    x^{j} \e^{N V_2( x)} ,
\ee
where the boundary terms are assumed to vanish every time due to the exponential terms. We use the following identity:
\be 
\dyi{r} \bigl[(ay)^s \e^{a xy}\bigr] = \dxi{s} \bigl[(ax)^r \e^{a xy}\bigr].
\ee 
Indeed, using Leibniz formula:
\begin{align}
\dyi{r} \left[(ay)^s \e^{a xy}\right]  = \sum_{k=0}^r \binom{r} {k} a^s\dyi{k}\left[ y^s \right]\dyi{r-k}\left[ \e^{a xy} \right]
 = \e^{a xy} \sum_{k=0}^{\min(r,s)} \frac 1 {k!} \frac {s!}{(s-k)!}\frac {r!}{(r-k)!} \frac{(ay)^{s} (ax)^{r} }{(axy)^k}.
\end{align}
The same expression is obtained developing $ \dxi{s} \left[(ax)^r \e^{a xy}\right]$. Using this identity,  \eqref{eq:point-proof-int-form} is equal to:
\be 
(-1)^{s} \int \d x\, \d y\, \e^{ NV_1(y)} \left[\dxi{s} \e^{- xy} x^r\right]   x^{j} \e^{N V_2( x)} =\int \d x\, \d y\, \e^{- xy} y^s    \e^{ NV_1(y)} \left[\dxi{s}x^{j} \e^{N V_2( x)}\right] ,
\ee
where the equality is obtained by doing again  $s$ integrations by part. 
We thus recover \eqref{eq:general-new-expression-p-s-int}.

\subsection{Potential application to new orthogonal polynomial method}
\subsubsection{Equivalence between the diagonalized differential formulations for one-matrix models}
\label{subsub:invert-diff-diag-gaussian}
 Performing the steps of Sec.~\ref{subsub:slater-diff-two-mat}, but starting from the differential formulation \eqref{eq:diff-diagonal-new-herm} of one-matrix models leads to:
\begin{equation}
\label{eq:Slater-diff-one-mat}
Z_{V}= f_N\, \det\left\{ \left[ \dxi{i} \e^{\frac 1 {2N} \dxi{2}} x^{j} \e^{N V( x)} \right]_{x=0} \right\}_{0\le i,j \le N-1}.
\end{equation}
This is also the differential formulation of the Slater determinant expression \eqref{eq:Slater-diff} for a two-matrix model with $V_1(x)=\frac{x^2}{2}$.

On the other hand, performing steps similar to Sec.~\ref{subsub:det-form-usual} to express \eqref{eq:diff-diagonal-usual-herm} in determinant form, we obtain: 
\begin{equation}
\label{eq:Slater-diff-one-mat-bis}
Z_{V}= N! e_N\, \det\left\{ \left[ \e^{\frac 1 {2N} \dxi{2}} x^{i+j} \e^{N V( x)} \right]_{x=0} \right\}_{0\le i,j \le N-1}.
\end{equation}

From the simpler formulation \eqref{eq:Slater-diff-one-mat-bis}, one can use \emph{orthogonal} polynomials satisfying
\be 
\label{eq:orthogonal-differential}
\left[  \e^{\frac 1 {2N} \dxi{2}} P_i(x)P_j(x) \e^{N V( x)}\right]_{x=0} = \delta_{ij} p_i
\ee
instead of \emph{biorthogonal} polynomials \eqref{eq:bil-form-int} satisfying 
\be
\label{eq:biorthogonal-differential}
\left[ P_i\Bigl(\dx\Bigr) \e^{\frac 1 {2N} \dxi{2}} Q_j(x) \e^{N V( x)} \right]_{x=0} = \delta_{ij} p_i,
\ee
{\it a priori} leading to a simpler resolution.

\

The computations of Sec.~\ref{sub:integration} in the differential formulation, in the case where $V_1(x)=\frac{x^2}{2}$ ($p=2$, $\alpha_2=1$), prove the equivalence between \eqref{eq:Slater-diff-one-mat} and \eqref{eq:Slater-diff-one-mat-bis}. Indeed for $p=2$, $r(i)=i$ so that the coefficients  $({N^{p-1}}/ {\alpha_p})^{r(i)}=N^i$ in the determinant just provide an overall factor $N^{N(N-1)/2}$, and   $f_NN^{\frac {N(N-1)} 2}= N! e_N$.

\

In integral form, the proof is simpler, as the Gaussian integration for $V_1(x)=\frac{x^2}{2}$ can be carried out explicitly after changing variables for  $x'=N x / \imath$.\footnote{There is an important subtlety in changing variables from a real variable to a pure complex variable: in addition to the Jacobian of the change of variables, an additional factor $-1$ has to be taken into account. This is detailed in Appendix~\ref{appendixB}, see \eqref{eq:change-variables-pure-complex}. The problem does not appear for the one-matrix model in the computations in differential formulation of Sec.~\ref{sub:integration}, or using the heat kernel \eqref{eq:heat-kernel}.} Equivalently, one may recover directly the Slater determinant form of the partition function of one-matrix models in the integral formulation, that is  
\begin{equation}
\label{eq:slater-1-mat-usual}
Z_{V_2}= \frac {N!}{b_N} \det\left\{ \int \d y\,  \e^{-N\frac {y^2}2}\e^{NV_2(y)}y^{i+j}  \right\}_{0\le i,j\le N-1},
\end{equation} 
from the differential formulation of the Slater determinant \eqref{eq:Slater-diff} (after the change of variables \eqref{eq:ex-p}),
\begin{equation}
\label{eq:Slater-diff-onemat}
Z_{{x^2}/2,V_2}= f_N \det\left\{ \left[ \dxi{i}  \e^{\frac 1 {2N} \dxi{2}} x^{j} \e^{N V_2( x)} \right]_{x=0} \right\}_{0\le i,j \le N-1},
\end{equation}
by using the \emph{heat kernel} formulation:
\be
\label{eq:heat-kernel}
\e^{t \dxi{2}} F(x) = \int \d y\, K_t(x-y) F(y), \qquad K_t(x-y) = \frac 1 {\sqrt{4\pi t}}e^{-\frac{(x-y)^2}{4t}}, 
\ee
for $t=\frac 1 {2N}$ and $F(x)= x^{j} \e^{N V_2(x)}$. We get: 
\be
\label{eq:C-second-onematrix}
Z_{{x^2}/2,V_2}= f_N \det\left\{ \sqrt{\frac N {2\pi}}  \int \d y \left[ \frac{\partial^{i}}{\partial x^{i}}   \e^{-\frac{N (x-y)^2}{2}}\right]_{x=0}   y^{j} \e^{N V_2( y)} \right\}_{0\le i,j \le N-1},
\ee
where the bracket evaluating $x$ at 0 has been moved inside the integral, as only the derivative of the kernel depends on $x$. One can verify that 
\[
\left[ \frac{\partial^{i}}{\partial x^{i}}   \e^{-\frac{N (x-y)^2}{2}}\right]_{x=0}  =N^{i} \tilde Q_i(y) \e^{-\frac{N}2 y^2},
\]
where $\tilde Q_i(y)$ is a monic polynomial of degree $i$, so we do recover  \eqref{eq:slater-1-mat-usual} as $f_N (\frac N {2\pi})^{\frac N 2} N^{\frac {N(N-1)}2} = N!/b_N$.

One can also use the heat kernel to find the integral representation of \eqref{eq:diff-diagonal-new-herm} and the one-matrix version of \eqref{eq:Slater-first-expr}, again obtained by setting $V_1(x) = \frac{x^2}{2}$. Applying the $N$-dimensional version of \eqref{eq:heat-kernel} to \eqref{eq:diff-diagonal-new-herm} for each $x_i$ we find
\be
Z_V = \left(\frac{N}{2\pi}\right)^{\frac{N}{2}} \left[ \frac{1}{\vd(\x)}  \int \d\y \, \e^{-\frac{N(\x-\y)^2}{2}} \vd(\y)\, e^{N V(\y)} \right]_{\x=0}.
\label{eq:diff-diagonal-new-herm-heat}
\ee
Similarly, in the one-matrix version of \eqref{eq:Slater-first-expr} we can use the heat kernel to find
\be
Z_{x^2/2,V_2} = \left[ \frac{1}{\vd(\x)} \det  \left\{ \frac{1}{\sqrt{2\pi}}\int \d y\, \e^{-\frac{(x_i-y)^2}{2}} y^{j-1} \e^{N V_2\left(\frac{1}{\sqrt{N}} y\right)} \right\}_{1\le i,j\le N} \right]_{\x=0}.
\label{eq:Slater-first-expr-heat}
\ee

\subsubsection{Application to new orthogonal polynomial methods}

As we have seen at the beginning of the present section, the computations of Sec.~\ref{sub:integration} in the case where $V_1(x)=\frac{x^2}{2}$, prove the equivalence in the differential formulation between \eqref{eq:Slater-diff-one-mat} and \eqref{eq:Slater-diff-one-mat-bis}, and the latter expression allows using orthogonal polynomials \eqref{eq:orthogonal-differential} instead of biorthogonal polynomials \eqref{eq:biorthogonal-differential}. One may naturally wonder if for potentials $V_1$ of degree higher than two, our new expressions \eqref{eq:general-new-expression-p-s-diff}, \eqref{eq:general-new-expression-p-s-int} could also allow the use of orthogonal polynomials instead of biorthogonal polynomials for two-matrix models.

\ 

To this aim, we reorganize the lines of the matrix $M$ in \eqref{eq:general-new-expression-p-s-diff} according to their remainder modulo $p-1$, that is:
\begin{equation}
\label{eq:new-determinant}
Z_{V_1, V_2} = h_N \det\left\{ \tilde M_{i,j} \right\}_{0\le i,j \le N-1}, \quad \mathrm{where}\quad 
\tilde M = \left(
\begin{array}{c}
 T_0 \\ \hline
 \vdots
 \\
 \hline
T_{p-1} 
\end{array}
\right) \quad \mathrm{and} \quad (T_s)_{r,j} = M_{(p-1)r + s\; ,\; j}. 
\end{equation}
where $h_N = f_N\, \mathrm{sgn} (N, p)$, $\mathrm{sgn} (N, p)$ being the parity of the permutation of rows, and we recall (\eqref{eq:general-new-expression-p-s-diff}) that in differential formulation,
\be
\label{eq:general-new-expression-p-s-recall}
M_{(p-1)r + s\; ,\; j} = ({N^{p-1}}/ {\alpha_p})^{r} \left[ \e^{NV_1(\dx)} x^r \dxi{s} \Bigl[x^{j} \e^{N V_2( x)} \Bigr]\right]_{x=0}.
\ee 
If $s_0$  and $r_0$ are respectively the remainder and the quotient of the Euclidean division of $N-1$ by $p-1$, then for $0\le s\le s_0$, the matrix $T_s$ has $r_0+1$ lines, while for $s_0< s < p-1$, $T_s$ has $r_0$ lines.  

\ 

By reorganizing the columns of $\tilde M$, and  the lines of $T_s$ independently for each $s$, we may replace $({N^{p-1}}/ {\alpha_p})^{-r}(T_s)_{r,j} = ({N^{p-1}}/ {\alpha_p})^{-r}M_{(p-1)r + s\; ,\; j}$ in the determinant by 
\be
\label{eq:new-determinant-ortho}
 \left[ \e^{NV_1(\frac 1 N \dx)} P_r^{(s)}(x) \dxi{s} \Bigl[Q_j(x) \e^{N V_2( x)} \Bigr]\right]_{x=0}  =  \int\frac{\d x\, \d y}{2\pi / \imath}\, \e^{-xy} \e^{NV_1(\frac 1 N y)} P_r^{(s)}(x) \dxi{s} \Bigl[Q_j(x) \e^{N V_2( x)} \Bigr],
\ee
where for $0\le s < p-1$, the $P_r^{(s)}(x)$ are monic polynomials of degree $r$, and $Q_j$ is a monic polynomial of degree $j$.

\ 

One may for instance choose $P_r^{(s)} = Q_r$ for all $0\le s < p-1$. We use the following notation for the determinant obtained this way:
\begin{equation}
\label{eq:new-determinant-ortho-fin}
Z_{V_1, V_2} = h_N \det\left\{ W_{i,j} \right\}_{0\le i,j \le N-1}, \quad \mathrm{where}\quad 
W = \left(
\begin{array}{c}
 J_0 \\ \hline
 \vdots
 \\
 \hline
J_{p-1} 
\end{array}
\right). 
\end{equation}
For $s=0$, the elements $(J_0)_{r,j}$ of the resulting matrix $J_0$ define the symmetric bilinear form 
\be 
\langle Q_r \mid\mid Q_j \rangle = \left[ \e^{NV_1(\frac 1 N \dx)} \e^{N V_2( x)} Q_r(x) Q_j(x)  \right]_{x=0} = \frac{\imath}{2\pi} \int\d x\,  Q_r(x) Q_j(x) \e^{N V_2( x)} \int \d y\, \e^{-xy} \e^{NV_1(\frac 1 N  y)},
\ee 
and requiring  $\langle Q_r \mid\mid Q_j \rangle = h_r \delta_{r,j}$ defines a family of orthogonal polynomials for the following weight: 
\be 
\label{eq:ortho-weight}
\e^{NV_1(\frac 1 N  \dx)}\e^{N V_2( x)} \quad \leftrightarrow\quad  \e^{N V_2( x)} \int \d y\, \e^{-xy} \e^{NV_1( \frac 1 N  y)}.
\ee
In addition to the usual three-terms recurrence $Q_{n+1}(x) = (x-\beta_n) Q_n - \frac{h_n}{h_{n-1}} Q_{n-1}$ satisfied by any family of orthogonal polynomials,  to obtain families of recurrence relations on the coefficients $\beta_n$ and $h_n$, one may for instance use the fact that 
\begin{align}
    \left[ \e^{NV_1(\frac 1 N \dx)} V_1'\Bigl(\frac 1 N\dx\Bigr)F(x) \right]_{x=0} & =     \left[ \left[ \e^{\dx \dy} V_1'\Bigl(\frac 1 Ny\Bigr)\e^{NV_1(\frac 1 N y)} \right]_{y=0}F(x)\right]_{x=0}\\&= \left[ \left[ \e^{\dx \dy} \dy\e^{NV_1(\frac 1 N y)} \right]_{y=0}F(x) \right]_{x=0} ,
\end{align} 
so that 
\be 
\label{eq:equation-E-ortho-pol}
\left[ \e^{NV_1(\frac 1 N\dx)} V_1'\Bigl(\frac 1 N\dx\Bigr)F(x) \right]_{x=0} = \left[  \e^{NV_1(\frac 1 N\dx)}  xF(x) \right]_{x=0},
\ee 
applied to $F(x) = Q_r(x) Q_j(x) \e^{N V_2( x)}$ for any $r,j$. For instance for the case $V_1(x) = x^3 / 3$, for which the weight for the orthogonal polynomials includes the Airy function as a factor, this equation 
leads to the family of equations 
\be 
\frac 2 N  \langle V_2' Q_r' \mid\mid Q_j \rangle + \frac 2 N  \langle V_2' Q_r \mid\mid Q_j' \rangle + \frac 1 {N^2}\Bigl(\langle Q_r'' \mid\mid Q_j \rangle + \langle Q_r \mid\mid Q_j'' \rangle + 2\langle  Q_r' \mid\mid Q_j' \rangle\Bigr) +  \langle R\, Q_r \mid\mid Q_j \rangle = 0,
\ee 
with $R(x) = \frac 1 N  V_2''(x) +  ( V_2'(x))^2 -  x$. 

\

The rest of the determinant \eqref{eq:new-determinant-ortho-fin} for $s>0$, obtained from \eqref{eq:new-determinant-ortho}, also involves sums of terms of the form $\langle g\, Q_r \mid\mid Q_j^{(k)} \rangle $ for some polynomials $g(x)$. Computing the determinant therefore requires knowledge about the derivatives of the families of orthogonal polynomials with weight \eqref{eq:ortho-weight}. It is not clear at this point whether the family of orthogonal polynomials can indeed be constructed and whether the rest of the determinant can be computed this way.

On the other hand, everything seems to simplify drastically at large $N$: for the choice $P_r^{(s)} = Q_r$ for all $0\le s < p-1$, the leading contributions in $N$ to the matrix elements \eqref{eq:new-determinant-ortho} for the  part of the determinant corresponding to $s>0$ are obtained when all $s$ derivatives act on $e^{NV_2(x)}$, raising a factor  $N^s(V_2'(x))^s$, so that the elements of the determinant simplify to
\be
\label{eq:new-determinant-ortho-lead}
(J_s)_{r,j} \sim_{N\rightarrow \infty} \frac{N^{(p-1)r + s}}{\alpha_p^r}\left \langle (V_2')^s Q_r \mid\mid Q_j \right\rangle .
\ee 
In the same way, the leading contribution to the family of relations \eqref{eq:equation-E-ortho-pol} seems to be given by 
\be 
\label{eq:equation-E-ortho-pol-largeN}
\left[ \e^{NV_1(\frac 1 N\dx)} e^{NV_2(x)}V_1'\bigl(V_2'(x)\bigr) Q_r(x)Q_j(x) \right]_{x=0} = \left[  \e^{NV_1(\frac 1 N\dx)}  xF(x) \right]_{x=0}, \quad \Leftrightarrow \quad \langle U Q_r \mid\mid Q_j\rangle =0,
\ee 
 with $U(x) = V_1'\bigl(V_2'(x)\bigr) - x$.
However, this already goes beyond the original scope of the paper, and we leave this for future work.

\section*{Acknowledgements}

The work of JB is supported by a Projectruimte grant of the Foundation for Fundamental Research on Matter (FOM, now defunct), financially supported by the Netherlands Organisation for Scientific Research (NWO).
The work of JT is funded by the German Research Foundation~(DFG) 
primarily under the project number 418838388 and further under Germany's Excellence Strategy EXC 2044–390685587, Mathematics M\"unster: Dynamics–Geometry–Structure. LL acknowledges support of the START-UP 2018 programme with project number 740.018.017, which is financed by the Dutch Research Council (NWO). LL thanks R.~Gurau and V.~Rivasseau for mentioning references using the differential formulation, and V.~Bonzom for pointing out the work of Orlov.

\appendix
\renewcommand\thesubsubsection{}
\titleformat{\subsubsection}[runin]{\bfseries}{\thesubsubsection}{0em}{}[.]
\section{Normalizations of the differential formulations}
\label{app:normalizations}

\subsubsection{Normalization of the differential formulation of diagonalized one-matrix models}
\label{par:norm-1}

The normalization $e_N$ in \eqref{eq:diff-diagonal-usual-herm} is computed developing the Vandermonde determinants as in Sec.~\ref{subsub:det-form-usual}:
\be
\begin{split}
\frac 1 {e_N} &= \left[ \e^{\frac 1 {2N}(\frac{\pd}{\pd \x})^2}\vd^{2}(\x)\right]_{\x=0} \\
&= N! \det_{i,j} \left\{ \left[ \e^{\frac 1 {2N} \dxi{2}} x^{i+j} \right]_{x=0} \right\}  \\
&= N! \det_{i,j}\Biggl\{ \sum_{n\ge 0}\frac {(2N)^{-n}} {n!} \left[ \dxi{2n} x^{i+j} \right]_{x=0} \Biggr\},
\end{split}
\ee
where the indices in the determinants range from 0 to $N-1$. The bracket in the rightmost expression is non-vanishing only if $2n=i+j$, in which case $\left[ \dxi{2n} x^{i+j} \right]_{x=0} = (i+j)!$. Therefore, 
$
1/ {e_N} = N!  \det(R) (2N)^{-N(N-1)/2},
$
where the $N\times N$ matrix $R$ is such that $R_{ij}=0$ for $i+j$ odd, and $R_{ij} = (i+j)!/[(i+j)/2]!$ for $i+j$ even. This determinant can be computed to be $\det(R) = 2^{N(N-1)/2} \prod_{j=1}^{N-1} j!$, so that
$
e_N = N^{N(N-1)/2} / \prod_{j=1}^N j!.
$

\subsubsection{Normalization of the differential formulation of diagonalized two-matrix models}
\label{par:norm-2}

The normalization is the same as for the one-matrix model, since:
\begin{align}
\left[ \e^{\frac 1 N \sum_{i=1 }^N\frac{\pd}{\pd a_i}\frac{\pd}{\pd b_i}}\,\Delta(\ba)\Delta(\bb) \right]_{\ba=\bb=0} 
& = N! \det_{i,j} \left\{ \left[ \e^{\frac 1 N \sum_{i=1 }^N\frac{\pd}{\pd a}\frac{\pd}{\pd b}}\, a^{i}b^{j} \right]_{a=b=0} \right\}  
\\ &= N! \det_{i,j}\Biggl\{ \sum_{n\ge 0}\frac {(N)^{-n}} {n!} \left[ \frac{d^{n}}{d a^{n}} a^i\right]_{a=0}\left[ \frac{d^{n}}{d b^{n}} b^j\right]_{b=0} \Biggr\}
\\ &= N! \det_{i,j}\Biggl\{ \delta_{i,j} i! N^{-i} \Biggr\} = \frac 1 {e_N}.
\end{align}

\section{Alternative representations 
using generalized heat kernels}
\label{appendixB}

In Sec.~\ref{subsub:invert-diff-diag-gaussian}, we have detailed how, for one-matrix models,  the Slater determinant form \eqref{eq:slater-1-mat-usual} in the integral formulation can be recovered directly from that in differential formulation \eqref{eq:Slater-diff-onemat} by using the heat kernel formula \eqref{eq:heat-kernel}.

In this appendix, we detail for completeness the analogous computation for two-matrix models, that is,  the integral analogues of expressions \eqref{eq:diff-diagonal-new}, \eqref{eq:Slater-first-expr}, and \eqref{eq:Slater-diff} using generalized heat kernels. Together these form the back lower edge of the diagram shown in Fig. \ref{fig:diagram-two-matrix}.

Our starting point is the differential representation \eqref{eq:diff-diagonal-new}:
\be
\label{eq:diff-app}
Z_{V_1,V_2}=  \left[ \frac 1 {\vd(\x)} \e^{N V_1( \frac 1 {\sqrt N} \frac{\pd}{\pd \x})} \vd(\x) \,\e^{NV_2(\frac 1 {\sqrt N}  \x)} \right]_{\x=0}
\, .\ee
In \eqref{eq:heat-kernel}, we used the heat kernel $K_t(x)$ to transform a differential representation of the one- model to an integral form. For two-matrix models, we can use the following analogous identity (see e.g. \cite{Robinson}):
\be
\label{eq:gen-heat-kernel}
 \e^{ W(\frac{d}{d \x})} F(\x) = \int_{\bR^N} d\y \,K_{W} (\x-\y) F(\y),
\ee
where $W$ is a polynomial, and $K_{W}$ is a generalized heat kernel with the following Fourier representation:
\be
K_{W} (\x-\y) = \frac {1} {(2\pi)^N} \int_{\bR^N} d\bp\,  \e^{\imath \bp \cdot (\x-\y) 
+ W(\imath \bp)}.
\ee
If we set $W(\x) = N V_1\left(\frac{1}{\sqrt{N}} \x\right)$ and $F(\x) = \vd(\x) e^{N V_2\left(\frac{1}{\sqrt{N}} \x\right)}$, we can substitute \eqref{eq:gen-heat-kernel} in \eqref{eq:diff-app} to find
\be
\label{eq:app-c1}
Z_{V_1,V_2}= \frac {1} {(2\pi)^N} \left[ \frac{1}{\vd(\x)} \int_{\bR^N} \d\y \int_{\bR^N} \d\bp\, \e^{\imath \bp \cdot(\x-\y)+NV_1(\frac 1 {\sqrt{N}}\imath \bp)} \vd(\y) \e^{N V_2(\frac 1 {\sqrt N}\y)}\right]_{\x=0}.
\ee
This is the integral analogue of \eqref{eq:diff-diagonal-new}. Subsequently, we can rewrite this expression as a Slater determinant by expanding the Vandermonde determinant, just like we did when rewriting \eqref{eq:diff-diagonal-new} to \eqref{eq:Slater-first-expr}, thus obtaining:
\be
\label{eq:app-c2}
Z_{V_1,V_2}= \frac {1} {(2\pi)^N}  \left[\frac{1}{\vd(\x)} \det\left\{ \int_{\bR} \d y \int_{\bR} \d p\,\e^{\imath p(x_i-y)+NV_1(\frac 1 {\sqrt{N}}\imath p)} y^{j} \e^{N V_2(\frac 1 {\sqrt N} y)}\right\}_{0\le i,j\le N-1} \right]_{\x=0}.
\ee
Finally, we can use identity \eqref{eq:identity-diff-det} to obtain
\be
\label{eq:app-c3}
Z_{V_1,V_2}= \frac {f_N} {(2\pi)^N}   \det\left\{ \int_{\bR} \d y   \int_{\bR} \d p\, \left[\frac{\partial^{i}}{\partial x^{i}}\e^{\imath p(x-y)}\right]_{x=0} \e^{NV_1(\frac 1 {\sqrt{N}}\imath p)} y^{j} \e^{N V_2(\frac 1 {\sqrt N} y)} \right\}_{0\le i,j\le N-1},
\ee
where the bracket evaluating $x$ at 0 could now be moved inside the determinant and the integral. This equation mirrors \eqref{eq:Slater-diff} using the generalized heat kernel. 
Carrying out the differentiation and evaluating $x$ at 0: 
\be
Z_{V_1,V_2}= \frac {f_N} {(2\pi)^N}   \det\left\{ \int_{\bR} \d y   \int_{\bR} \d p\, (\imath p)^i\e^{-\imath py} \e^{NV_1(\frac 1 {\sqrt{N}}\imath p)} y^{j} \e^{N V_2(\frac 1 {\sqrt N} y)} \right\}_{0\le i,j\le N-1}.
\ee
To verify that the equation is well-normalized we may use \eqref{eq:Gaussian-mixt} to carry out the integration of the matrix elements. There is a subtlety however: \eqref{eq:Gaussian-mixt-0} and \eqref{eq:Gaussian-mixt}  have to be modified for $\alpha\in \imath \mathbb{R}$.
For $n=m=0$ and  $\alpha\in \imath \mathbb{R}$, \eqref{eq:Gaussian-mixt-0} should be replaced with:
\be
\label{eq:Gaussian-mixt-compl-0}
\int_{\bR^2}\d x\, \d y\, \e^{- \imath \alpha  N  xy } = \frac{2\pi} {\alpha N},
\ee
and for positive $n$ or $m$:
\be 
\int_{\bR^2}\d x\, \d y\, \e^{- \imath \alpha N  xy }\, x^n y^m = \frac {\delta_{n,m}  }{(-\imath N)^n} \frac{\partial^{n}}{\partial \alpha^{n}} \int_{\bR^2}\d x\, \d y\, \e^{- \imath \alpha N  xy } =  \frac {\delta_{n,m}  } {(-\imath N)^n} \frac{\partial^{n}}{\partial \alpha^{n}} \frac{2\pi}{\alpha N} 
,
\ee
so \eqref{eq:Gaussian-mixt} has to be modified for:
\be
\label{eq:Gaussian-mixt-compl}
\int_{\bR^2}\d x\, \d y\, \e^{- \imath \alpha N  xy }\, x^n y^m = \delta_{n,m}\,  2\pi \imath \,  \left(\frac 1 {\imath \alpha N}\right)^{n+1} n!.
\ee
These expressions allow showing that 
\be 
\label{eq:change-variables-pure-complex}
\int_{\bR^2} \d y\, \d p  \, \e^{-\imath py}  (\imath p)^i\e^{NV_1(\frac 1 {\sqrt{N}}\imath p)} y^{j} \e^{N V_2(\frac 1 {\sqrt N} y)} = \imath  \int_{\bR^2} \d y\, \d x  \, \e^{-xy} x^i  \e^{NV_1(\frac 1 {\sqrt{N}}x)} y^{j} \e^{N V_2(\frac 1 {\sqrt N} y)},
\ee
in the sense that their perturbative expansions match, where \eqref{eq:Gaussian-mixt-compl} has to be used to expand the left hand side and \eqref{eq:Gaussian-mixt} for the right hand side, which explains the factor $\imath$, while $\imath^{-1}$ would naively be expected as a Jacobian by directly changing variables for $x=\imath p$ in the integral. Finally, changing variables for $\sqrt N x'=x$ and $\sqrt N y'=y$, we  recover the integral Slater determinant form \eqref{eq:slater-2-mat-usual}, since $\frac{f_N}{(2\pi)^N} N^{N(N+1)/2} \imath^N  = \frac 1 {d_N}$.

\bibliographystyle{JHEP}
\bibliography{mmdiff}

\end{document}